\documentclass[aps,prl,twocolumn,noshowpacs]{revtex4-1}
\usepackage{graphicx,epsfig,epstopdf,subfigure}
\usepackage{float}
\usepackage{amsbsy}
\usepackage{amssymb}
\usepackage{amsmath}
\usepackage{grffile}
\usepackage{xifthen}
\usepackage{longtable,tabu,tabularx,multirow}

\graphicspath{{Figures//}}

\date{\today}

\begin{document}

\newcommand{\eqnref}[1]{Eq.~\ref{#1}}
\newcommand{\figref}[2][]{Fig.~\ref{#2}#1}
\newcommand{\citeref}[1]{Ref. \cite{#1}}
\newcommand{\tableref}[1]{Table~\ref{#1}}

\title{
Kibble-Zurek Mechanism in a Spin-1 Bose-Einstein Condensate}
\author{
M. Anquez, B.A. Robbins, H.M. Bharath, M.J. Boguslawski, T.M. Hoang, and M.S. Chapman}

\affiliation{School of Physics, Georgia Institute of Technology,
  Atlanta, GA 30332}

\begin{abstract}

We observe power-law scaling of the temporal onset of excitations with quench speed in the neighborhood of the quantum phase transition between the polar and broken-axisymmetry phases in a small spin-1 ferromagnetic Bose-Einstein condensate. 
As the system is driven through the quantum critical point by tuning the Hamiltonian, the vanishing energy gap between the ground state and first excited state causes the reaction time scale of the system to diverge, preventing it from adiabatically following the ground state. 
We measure the temporal evolution of the spin populations for different quench speeds and determine the exponents characterizing the scaling of the onset of excitations, which are in good agreement with the predictions of the Kibble-Zurek mechanism. 

\end{abstract}

\maketitle

In a second-order (or continuous) quantum phase transition (QPT), a qualitative change in the system's ground state occurs at zero temperature when a parameter in the Hamiltonian is varied across a quantum critical point (QCP)  \cite{Sachdev99}. Near the critical point of the transition, the time scale characterizing the dynamics of a system diverges, and the scaling of this divergence with respect to the quench speed through the phase transition is characterized by universal critical exponents.
The Kibble-Zurek mechanism (KZM) as originally formulated characterizes the formation of topological defects when a system undergoes a continuous phase transition at a finite rate. 
This concept was first conceived by Kibble in his study on topology of cosmic domains and strings in the early universe \cite{Kibble76, Kibble80}, and it was later extended by Zurek \cite{Zurek85, Zurek93, Zurek96} who suggested applying these symmetry breaking ideas to condensed matter systems, such as superconductors and superfluids.

This seminal work was followed by many theoretical studies applying the KZM to cosmology, condensed matter, cold atoms and more \cite{Laguna97, Laguna98, Yates98, Anglin99, Dziarmaga99, Antunes99, Stephens99, Stephens02, Zurek05, Dziarmaga05, Polkovnikov05, Damski06, Cherng06, Lamacraft07, Damski07, Fubini07, Zurek09, Dziarmaga10}.
In parallel, the KZM has been studied experimentally and verified in a large variety of systems, including liquid crystals \cite{Chuang91,Bowick94}, $^{4}$He \cite{Hendry94} and $^{3}$He \cite{Ruutu96,Bauerle96}, optical Kerr media \cite{Ducci99}, Josephson junctions \cite{Carmi00}, superconducting films \cite{Maniv03}, and annealed glass \cite{Monaco06}. 
There has also been significant interest in the KZM in the cold atom community. In recent years, it has been observed in ion chains \cite{delCampo10,Pyka13,Ejtemaee13,Ulm13}, in atomic gases in optical lattices \cite{Chen11}, and in Bose-Einstein condensates (BECs), through the formation of spatial domains during condensation \cite{Donner07,Navon15}, vortices \cite{Scherer07,Weiler08}, creation of solitons \cite{Lamporesi13} and supercurrents \cite{Corman14}.
Only a few experiments have explored the KZM using QPTs (i.e. at  zero temperature), namely an investigation of the Mott insulator to superfluid transition \cite{Braun15} and, in a recent preprint \cite{Cui15}, an ion chain cooled to the ground state. There has been related work investigating universal scaling in optical lattices \cite{Zhang12} and recently in the miscible-immiscible transition in a two-component Bose gas \cite{Nicklas15}.

\begin{figure}[b]
	\begin{center}
	\begin{minipage}{3.25in}
		 \includegraphics[scale=1]{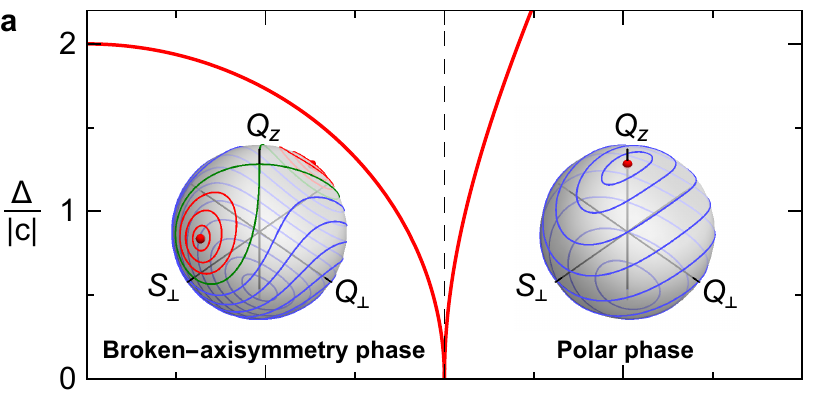}	 
		 \includegraphics[scale=1]{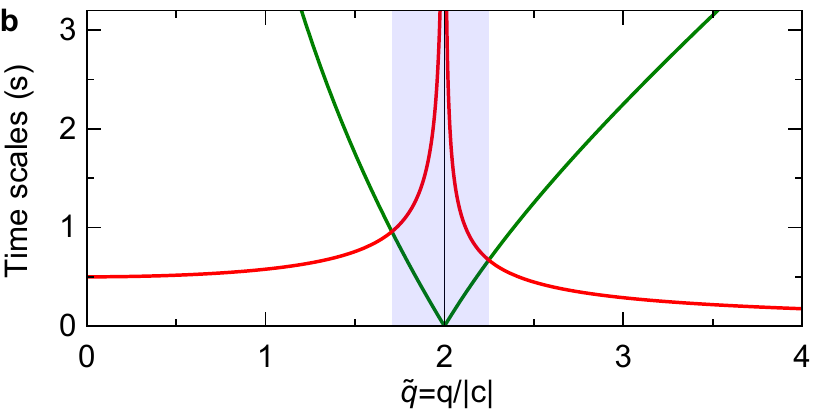}	 
	\end{minipage}
	\end{center}
    \caption{\textbf{Concept}. (a) The energy gap $\Delta$ between the ground state and first excited state is plotted as a function of the quadratic Zeeman energy $q$ (in units of $|c|$).  The gap vanishes at the critical point $q_c = 2|c|$, shown by the vertical dashed line. The spheres show the spin-nematic phase space for different values of $q$: (left) broken-axisymmetry phase ($q < 2|c|$) and (right) polar phase ($q > 2|c|$). The ground state in each phase is indicated with red dots. (b) The `freeze-out' region for a given ramp speed (blue shaded) is determined by the intersection of the  minimum response time (red) and the effective speed of the ramp (green).  See text for details.
    }
\label{fig:energyGapPlot}
\end{figure}

A ferromagnetic spin-1 ($^{87}$Rb) BEC exhibits a QPT between a symmetric polar (P) phase and a broken-axisymmetry (BA) phase \cite{Murata07,Lamacraft07} due to the competition between magnetic and collisional spin interaction energies. There have been several theoretical studies predicting KZM power-law scaling of the spin excitations for slow quenches through the critical point \cite{Damski07, Damski08, Lamacraft07, Saito07, Saito13, Damski09}. 
In this Letter, we investigate such slow quenches in small spin-1 $^{87}$Rb condensates and demonstrate power-law scaling in the onset of spin excitations as a function of the quench speed from the polar to BA phases. A distinguishing feature of our KZM investigation is that the excitations are not manifest as spatial defects. Our small spin condensates are in the single mode approximation where all the atoms share the same spatial wave function, so unlike in spatially extended systems where the KZM is manifest in topological defects, the excitations appear in the temporal evolution of the spin populations.
In this study, we investigate slow quenches through the critical point, which distinguishes this work from several previous investigations of sudden quenches  \cite{Sadler06, Hamley12, Gerving12, Hoang13}.

\begin{figure*}[t!]
	\begin{center}
	\includegraphics[width=3.25in]{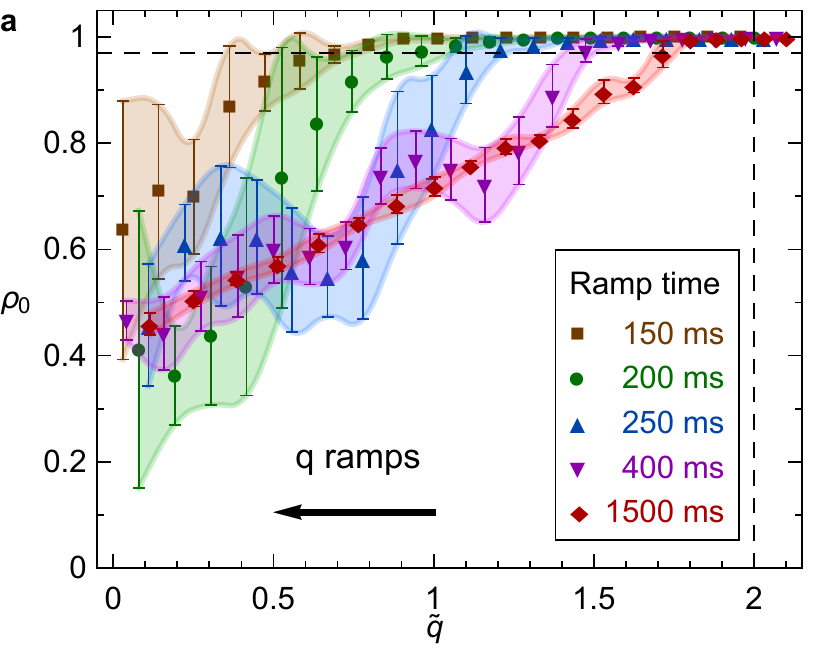}	
\hfill
	\includegraphics[width=3.25in]{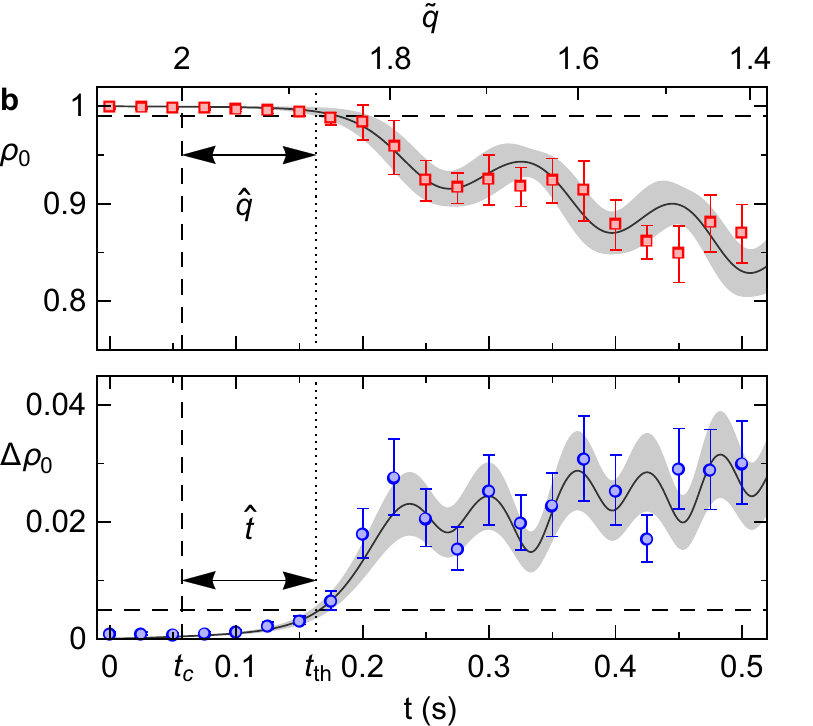}
	\end{center}
\caption{\textbf{Quench dynamics.} (a) Measurements of $\rho_0$ for different ramp times as a function of $\tilde{q} = q/|c|$. The ramp times shown correspond to the duration to ramp the magnetic field from $\tilde{q} = 2.2$ to $\tilde{q}=0$. The longer ramps show evolution after a smaller change in $\tilde{q}$ than the shorter ramps. For the latter, the system stay frozen in the polar phase ground state ($\rho_0=1$) until a larger change in $\tilde{q}$, as expected from the KZM. The horizontal and vertical dashed line indicate the $\rho_0$ threshold and the critical point respectively. (b) Measurements of $\rho_0$ (red squares) and its standard deviation $\Delta \rho_0$ (blue circles) during a typical experimental run where the magnetic field is slowly ramped down through the critical point such that the $q$ decreases linearly. The thresholds used to determine when the system `unfreezes' are shown as a horizontal dashed lines. The system shows good agreement with simulations (gray curves and envelopes showing $\pm$ one standard deviation) for long evolution times beyond the freeze-out period.  The top axis shows $\tilde{q}$, with the vertical dashed line at $\tilde{q} = 2$ marking the critical point. The dotted line indicates when the system crosses the determined threshold. }
\label{fig:rampsAndConcept}
\end{figure*}

The spin dynamics of our system in a homogeneous magnetic field (along the $z$-axis) can be described using the following Hamiltonian: 
$ \hat{H} = \tilde{c} \hat{S}^2  - \frac{q}{2} \hat{Q}_{z}$. The first term describes the spin interactions, where $\tilde{c}$ is the spin-dependent elastic collision coefficient related to the $s$-wave scattering lengths for collisions between pairs of atoms ($\tilde{c} < 0$ for ferromagnetic condensates), and $\hat{S}^2=\hat{S}_x^2+\hat{S}_y^2+\hat{S}_z^2$ is the total spin vector operator. 
The second term describes the quadratic Zeeman energy per particle. 
$\hat{Q}_z$ is proportional to the spin-1 quadrupole moment, $\hat{Q}_{zz}$ \cite{Hamley12}, and $q = q_z B^2$ is the quadratic Zeeman energy per particle, where $B$ is the magnitude of the magnetic field and $q_z \approx 71.6~\mathrm{Hz/G^2}$ (hereafter $h=1$).
The experiment is initialized with no longitudinal magnetization, $ \langle \hat{S}_z \rangle =0$, which is conserved by the Hamiltonian, so the linear Zeeman energy $p \hat{S}_z$ with $p \propto B $  can be ignored. 
In terms of the mean-field expectation values, the spin energy is:
\begin{equation}
H=\frac{c}{2} S^2 - \frac{q}{2} Q_{z}
\end{equation}
where $c \propto \tilde{c}$ is the spinor dynamical rate and $Q_z = 2\rho_0-1$, with $\rho_0$ being the fractional population in the $| F=1,m_F=0\rangle$ state. The states of the system can be represented on the $\{S_\perp, Q_\perp, Q_{z}\}$ unit sphere, where $S_\perp^2 = S_x^2 + S_y^2$ and $Q_\perp^2 = Q_{xz}^2 + Q_{yz}^2$ \cite{Hamley12}, as shown in \figref{fig:energyGapPlot}.

The quantum critical point between the P and BA phases occurs at $q_c = 2|c|$.  Using Bogoliubov theory, it can be shown that the BA phase ($q<q_c$) ground state has three excitation modes \cite{Murata07}. Two are gapless modes (in the long wavelength limit), which arise from the SO(2) symmetry breaking as predicted by the Goldstone theorem \cite{Goldstone61}, but the third mode has a non-zero eigenvalue, corresponding to the energy gap between the ground and first excited state (see \figref{fig:energyGapPlot}a).
\begin{equation}
\Delta =  \sqrt{q_c^2-q^2} \sim \bigl| q_c - q \bigl| ^{1/2}
\label{eqn:energyGap}
\end{equation}
where the approximation is valid near $q=q_c $. 

A universal feature of QPTs is that close to the critical point, the properties of the system are uniquely described by  critical exponents determining the functional form of the energy gap as a function of the parameters of the Hamiltonian: $\Delta \sim |g_c-g|^{z\nu} $, where $z$, $\nu$ are the critical exponents, $g$ is the tuning parameter, and $g_c$ is the critical point of the system \cite{Sachdev99}.  Comparing to Eq.~2 shows that $z\nu = 1/2$ for the spin-1 system.

Because the energy gap $\Delta$ vanishes at the critical point (ignoring finite-size effects), the system cannot cross the critical point adiabatically. The utility of the KZM is that it provides a universal prescription for quantifying the dynamical excitation based on the exponents $z,\nu$ that govern the equilibrium behavior of the system \cite{Zurek85,Damski05,Zurek05,Damski07}.
As illustrated in \figref{fig:energyGapPlot}b, two characteristic timescales can be compared to explain the behavior of the system initialized in the  ground state of one phase as it is driven across the QCP. 
The first is the reaction time of the system to changes in the Hamiltonian, which is inversely proportional to the energy gap $\Delta$. The second is $\Delta/\dot{\Delta}$ and describes how fast the system is driven through the critical point.
Close to the critical point, the reaction time is too large for the evolution to be adiabatic, and the evolution shifts to an impulse regime where the system remains frozen in the polar ground state.
When the two timescales become comparable again, the system unfreezes and is now in an excited state. The dynamics resume and the system is able to adiabatically evolve towards the BA ground state.

The freeze out time $\hat{t}$ between the crossing of the critical point and the recovery of adiabatic evolution is a function of the ramp speed and can be found from $ \Delta^{-1}(\hat{t}) =  \Delta/\dot{\Delta} |_{t=\hat{t}} $.
For the case of linear ramps of the control parameter of the Hamiltonian ($q$, in our case) in a quench time $\tau_Q$, then $\dot{q} \propto \tau_Q^{-1}$ and the power-law relation $\hat{t} \propto \tau_Q ^{\nu z/(1+\nu z)} = \tau_Q ^{\alpha}$, where $\alpha = 1/3$, is obtained (see Supplemental Information). Introducing the dimensionless ratio 
${\tilde{q}=q/|c|}$ 
and defining $\hat{q}$ as the change in $\tilde{q}$ between the crossing of the critical point and the resuming of dynamics, a power law scaling can also be derived for $\hat{q}$ using the same approach as for $\hat{t}$, which results in $\hat{q} \sim \tilde{\tau}_Q ^\frac{-1}{1+\nu z} = \tilde{\tau}_Q ^{\beta} $, where $\beta = -2/3$ and $\tilde{\tau}_Q$ is the inverse of the rate of change of $\tilde{q}$ at the critical point.

The experiments use small spin-1 $^{87}$Rb atomic Bose condensates in the $F=1$ hyperfine state confined in an optical dipole trap.  The condensates are initialized in the $m_F=0$ state in a high magnetic field, which is the polar ground state (see \figref{fig:energyGapPlot}a, right). To measure the power-law scaling for $\hat{t}$ and $\hat{q}$, the condensates are quenched across the QCP at different speeds, and the onset time (and corresponding value of $q$) for excitations from the polar ground state are determined from the time evolution of the mean value spin population $\rho_0$ and the fluctuation $\Delta \rho_0$.  In \figref{fig:rampsAndConcept}a, the results from several ramps are shown.  For each these ramps, the field is first lowered quickly to $q_0=2.2|c|$, and then ramped according to  $q(t) = q_0 - t/\tau_Q $ for a range of $\tau_Q$ values. 
For asymptotically slow ramps, the population $\rho_0$ should adiabatically follow the ground state value $ \rho_{0,\mathrm{GS}}=1/2+q/4|c|$ for $q<q_c=2|c|$.  From the data in \figref{fig:rampsAndConcept}a, it is clear that the population lags the ground state value by an amount that increasing for faster ramps, indicating the non-adiabatic crossing of the QCP.  

The determinations of $\hat{t}$ and $\hat{q}$ are shown in \figref{fig:rampsAndConcept}b, which show both $\rho(t)$ and $\Delta \rho_0(t)$ for a typical quench.  To determine when the system `unfreezes,' thresholds of $\rho_0 = 0.99$ and $\Delta\rho_0 = 0.005$ are used. As pointed out in \cite{Damski07}, the exponent is insensitive to the choice of the exact thresholds.
The freeze-out time is $\hat{t} = t_{\mathrm{th}} - t_c$, where $t_c$ is the time the system crosses the critical point and   $t_{\mathrm{th}}$ is where $\rho_0$ and $\Delta\rho_0$ reach their respective thresholds. $\hat{q}$ is determined similarly to $\hat{t}$, and is given by $\hat{q} = \tilde{q}(t_{\mathrm{th}}) - \tilde{q}(t_c)$.  
The use of $\tilde{q}$ allows us to incorporate the effect of the finite lifetime of the condensate ($\sim 2$~s) in the data analysis. 
The value $q/|c|$ is affected by the reduction of  density due to the finite lifetime of the condensate, as the spin interaction energy depends on the density and atom number as $c(t) \propto n(t) \propto N(t)^{2/5}$ (see Supplemental Information), which we account for by using $\tilde{q}=q(t)/|c(t)|$.

In order to be able to extract accurate values for $\hat{t}$ and $\hat{q}$, it is necessary to make a careful determination of $q_{c}=2|c|$. This is achieved by preparing the system in the polar ground state and  measuring the onset of fluctuations in $\rho_0$ following a fast quench from high field to a final field value in the neighborhood of the critical field, $B_c$. For final field values above $B_c$ the subsequent fluctuations are negligible, but there is a sharp onset of fluctuations below $B_c$. Using this approach, $B_c$ is determined with an uncertainty as low as 2~mG (Supplemental Information). This measurement of $B_c$ yields a value of $c$ found using $|c| = q_c/2 = q_z B_c^2/2$. Depending on the trap geometry, $c$ ranges from $-2.4$~Hz to $-8.2$~Hz.

\begin{figure}[h]
	\begin{center}
	\begin{minipage}{3.25in}
		 \includegraphics{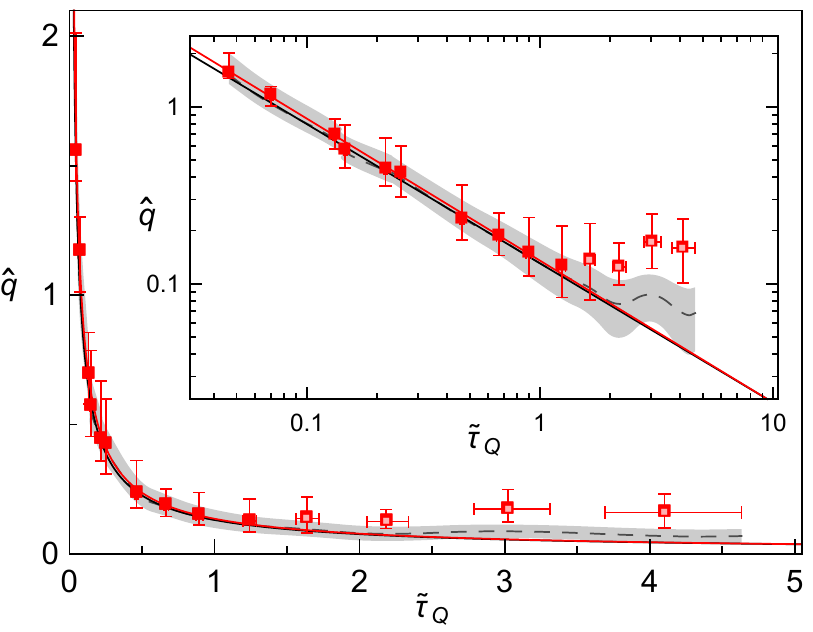}	
	\end{minipage}
	\end{center}
    \caption{\textbf{Power law fit}. $\hat{q}$ data (red squares) and simulations (dashed line with $\pm$ one standard deviation envelope) plotted with respect to the characteristic ramp time $\tilde{\tau}_Q$. The red and black solid lines correspond to fits to the data and simulations respectively. 
    The inset shows a linear fit to the log of the data for $\tilde{\tau}_Q < 1.5$ (solid markers), yielding a scaling exponent of $-0.80(8)$.
    }
\label{fig:qHatRho0LogPlot}
\end{figure}

\begin{figure}[h]
	\begin{center}
	\begin{minipage}{3.25in}
		 \includegraphics{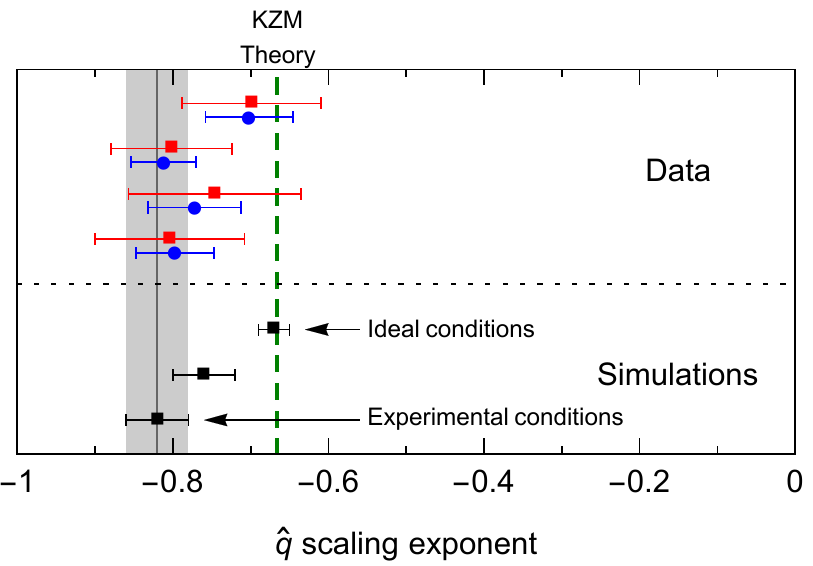}	
	\end{minipage}
	\end{center}
    \caption{\textbf{Summary of scaling exponents}. The red squares and blue circles indicate the scaling exponents for data sets analyzed using $\rho_0$ and $\Delta\rho_0$ as thresholds, respectively.  The results from \figref{fig:qHatRho0LogPlot} correspond to the second data set. 
    Simulations are represented by black squares, showing from top to bottom: ideal conditions, fast ramps starting at a high magnetic field, and experimental conditions. The latter is also corresponds to the gray line and shaded region.
    The dashed vertical line indicated the value predicted by the KZM theory. 
    }
\label{fig:partialSummaryPlot}
\end{figure}

In order to determine the scaling of the excitations as a function of ramp speed, the values of  $\hat{q}$ are plotted versus $\tilde{\tau}_Q$ in \figref{fig:qHatRho0LogPlot}.  The data are fit to a power law, which reveals good  agreement, except for the slowest ramps ($\tilde{\tau}_Q > 1.5$), which start to deviate from the fit.  This is likely due to the large amount of atom loss in this regime.
The inset in \figref{fig:qHatRho0LogPlot} shows the same data in a log-log plot along with a linear fit of the data between $ 0.04 < \tilde{\tau}_Q < 1.5$.  This fit yields the power law exponent $\beta = -0.80(8) $.
The data are compared with simulations matching the experiment conditions (gray shaded region) and the agreement is satisfactory.  In particular the power law exponent determined from the simulations is $\beta = -0.79(7)$.
The experiment was repeated multiple times over several months, and the results are  summarized in \figref{fig:partialSummaryPlot}.  The scaling exponents were determined from analyzing both  $\rho_0$ and $\Delta\rho_0$; the latter are shown as blue markers (see Supplemental information for results from all data sets).
 The fourth data set used a different trap geometry---an elongated cigar-shaped trap, which benefits from a longer lifetime of $\sim 15$~s. Even though the condensate was no longer in the single mode approximation, no spin domains were detected before the system crossed the thresholds used to determine the freeze-out time, and the measured $\beta = -0.80(10)$ is also in good agreement with simulations using the experimental settings.

The observed scaling exponents are self-consistent (within experimental uncertainty) and agree well with the simulations.  They are however, slightly more  negative than the $-2/3$ value derived above.  To investigate this discrepancy, we have performed simulations varying a wide range of parameters including atom number, ramp speeds, initial magnetic fields and condensate lifetime (Supplemental Information). 

Simulations performed in ideal conditions (infinite condensate lifetime, high initial magnetic field, and large number of atoms) yield $\beta = -0.67(2)$ for very long ramps ($\tilde{\tau}_Q > 2$), in excellent agreement with the value of $-2/3$ predicted by the KZM (see \figref{fig:partialSummaryPlot}).  However, for the faster ramps that we can measure due to the finite lifetime of the system, the simulations of the ideal case yield a more negative result with $\beta = -0.76(4)$.  The restriction of the simple theory to the slowest ramps was pointed out in \cite{Damski07}.
A second consequence of the lifetime of the condensate is that it prevents starting the magnetic field ramps at a value much higher than the critical point, since a large number or atoms would be lost by the time the system crossed the critical point. A compromise is reached experimentally by starting with a fast drop from a high field $(q = 17.1q_c)$ to a field closer to the critical point $(q = 1.1q_c)$, followed by slower ramps through the critical point. When we include this experimental step in the simulations, we get $\beta = -0.82(4)$, which agrees with the experimental results. From our simulations, the effects of atom loss are not important in the range of ramps that are used to determine the scaling exponent.

In summary, we have observed the Kibble-Zurek mechanism in spin-1 BEC by measuring the exciations as a function of the quench speed across the quantum critical point.  This is one of the few demonstrations of the KZM in a quantum phase transition. 
In the future, an investigation of the scaling of spin domain formation following slow quenches in larger condensates is a natural continuation of this study.
We are also interested in exploring modifications to the KZM in the spin-1 system due to finite size (quantum) effects. 
We have performed simulations for different numbers of atoms (holding all other parameters constant) that indicate that these effects should be observable in our experiments. Finally, understanding the behavior of quantum systems undergoing a QPT is relevant in the prospect of implementing adiabatic quantum computing \cite{Farhi01}.

\section{Acknowledgments}

We would like to thank Christopher Hamley and Benjamin Land for their early contributions to the theory and simulations, and we acknowledge support from the National Science Foundation (PHY--1208828).

\normalsize\normalfont

\bibliography{Anquez_KZM_refs}

\begin{thebibliography}{62}%
\makeatletter
\providecommand \@ifxundefined [1]{%
 \@ifx{#1\undefined}
}%
\providecommand \@ifnum [1]{%
 \ifnum #1\expandafter \@firstoftwo
 \else \expandafter \@secondoftwo
 \fi
}%
\providecommand \@ifx [1]{%
 \ifx #1\expandafter \@firstoftwo
 \else \expandafter \@secondoftwo
 \fi
}%
\providecommand \natexlab [1]{#1}%
\providecommand \enquote  [1]{``#1''}%
\providecommand \bibnamefont  [1]{#1}%
\providecommand \bibfnamefont [1]{#1}%
\providecommand \citenamefont [1]{#1}%
\providecommand \href@noop [0]{\@secondoftwo}%
\providecommand \href [0]{\begingroup \@sanitize@url \@href}%
\providecommand \@href[1]{\@@startlink{#1}\@@href}%
\providecommand \@@href[1]{\endgroup#1\@@endlink}%
\providecommand \@sanitize@url [0]{\catcode `\\12\catcode `\$12\catcode
  `\&12\catcode `\#12\catcode `\^12\catcode `\_12\catcode `\%12\relax}%
\providecommand \@@startlink[1]{}%
\providecommand \@@endlink[0]{}%
\providecommand \url  [0]{\begingroup\@sanitize@url \@url }%
\providecommand \@url [1]{\endgroup\@href {#1}{\urlprefix }}%
\providecommand \urlprefix  [0]{URL }%
\providecommand \Eprint [0]{\href }%
\providecommand \doibase [0]{http://dx.doi.org/}%
\providecommand \selectlanguage [0]{\@gobble}%
\providecommand \bibinfo  [0]{\@secondoftwo}%
\providecommand \bibfield  [0]{\@secondoftwo}%
\providecommand \translation [1]{[#1]}%
\providecommand \BibitemOpen [0]{}%
\providecommand \bibitemStop [0]{}%
\providecommand \bibitemNoStop [0]{.\EOS\space}%
\providecommand \EOS [0]{\spacefactor3000\relax}%
\providecommand \BibitemShut  [1]{\csname bibitem#1\endcsname}%
\let\auto@bib@innerbib\@empty
\bibitem [{\citenamefont {Sachdev}(1999)}]{Sachdev99}%
  \BibitemOpen
  \bibfield  {author} {\bibinfo {author} {\bibfnamefont {S.}~\bibnamefont
  {Sachdev}},\ }\href@noop {} {\emph {\bibinfo {title} {Quantum Phase
  Transitions}}}\ (\bibinfo  {publisher} {Cambridge University Press},\
  \bibinfo {address} {Cambridge, UK},\ \bibinfo {year} {1999})\BibitemShut
  {NoStop}%
\bibitem [{\citenamefont {Kibble}(1976)}]{Kibble76}%
  \BibitemOpen
  \bibfield  {author} {\bibinfo {author} {\bibfnamefont {T.}~\bibnamefont
  {Kibble}},\ }\href {\doibase doi:10.1088/0305-4470/9/8/029} {\bibfield
  {journal} {\bibinfo  {journal} {J. Phys. A}\ }\textbf {\bibinfo {volume}
  {9}},\ \bibinfo {pages} {1387} (\bibinfo {year} {1976})}\BibitemShut
  {NoStop}%
\bibitem [{\citenamefont {Kibble}(1980)}]{Kibble80}%
  \BibitemOpen
  \bibfield  {author} {\bibinfo {author} {\bibfnamefont {T.~W.}\ \bibnamefont
  {Kibble}},\ }\href@noop {} {\bibfield  {journal} {\bibinfo  {journal} {Phys.
  Rep.}\ }\textbf {\bibinfo {volume} {67}},\ \bibinfo {pages} {183} (\bibinfo
  {year} {1980})}\BibitemShut {NoStop}%
\bibitem [{\citenamefont {Zurek}(1985)}]{Zurek85}%
  \BibitemOpen
  \bibfield  {author} {\bibinfo {author} {\bibfnamefont {W.}~\bibnamefont
  {Zurek}},\ }\href@noop {} {\bibfield  {journal} {\bibinfo  {journal}
  {Nature}\ }\textbf {\bibinfo {volume} {317}},\ \bibinfo {pages} {505}
  (\bibinfo {year} {1985})}\BibitemShut {NoStop}%
\bibitem [{\citenamefont {Zurek}(1993)}]{Zurek93}%
  \BibitemOpen
  \bibfield  {author} {\bibinfo {author} {\bibfnamefont {W.}~\bibnamefont
  {Zurek}},\ }\href@noop {} {\bibfield  {journal} {\bibinfo  {journal} {Acta.
  Phys. Pol. B}\ }\textbf {\bibinfo {volume} {24}},\ \bibinfo {pages} {1301}
  (\bibinfo {year} {1993})}\BibitemShut {NoStop}%
\bibitem [{\citenamefont {Zurek}(1996)}]{Zurek96}%
  \BibitemOpen
  \bibfield  {author} {\bibinfo {author} {\bibfnamefont {W.~H.}\ \bibnamefont
  {Zurek}},\ }\href@noop {} {\bibfield  {journal} {\bibinfo  {journal} {Phys.
  Rep.}\ }\textbf {\bibinfo {volume} {276}},\ \bibinfo {pages} {177} (\bibinfo
  {year} {1996})}\BibitemShut {NoStop}%
\bibitem [{\citenamefont {Laguna}\ and\ \citenamefont
  {Zurek}(1997)}]{Laguna97}%
  \BibitemOpen
  \bibfield  {author} {\bibinfo {author} {\bibfnamefont {P.}~\bibnamefont
  {Laguna}}\ and\ \bibinfo {author} {\bibfnamefont {W.~H.}\ \bibnamefont
  {Zurek}},\ }\href@noop {} {\bibfield  {journal} {\bibinfo  {journal} {Phys.
  Rev. Lett.}\ }\textbf {\bibinfo {volume} {78}},\ \bibinfo {pages} {2519}
  (\bibinfo {year} {1997})}\BibitemShut {NoStop}%
\bibitem [{\citenamefont {Laguna}\ and\ \citenamefont
  {Zurek}(1998)}]{Laguna98}%
  \BibitemOpen
  \bibfield  {author} {\bibinfo {author} {\bibfnamefont {P.}~\bibnamefont
  {Laguna}}\ and\ \bibinfo {author} {\bibfnamefont {W.~H.}\ \bibnamefont
  {Zurek}},\ }\href@noop {} {\bibfield  {journal} {\bibinfo  {journal} {Phys.
  Rev. D}\ }\textbf {\bibinfo {volume} {58}},\ \bibinfo {pages} {085021}
  (\bibinfo {year} {1998})}\BibitemShut {NoStop}%
\bibitem [{\citenamefont {Yates}\ and\ \citenamefont {Zurek}(1998)}]{Yates98}%
  \BibitemOpen
  \bibfield  {author} {\bibinfo {author} {\bibfnamefont {A.}~\bibnamefont
  {Yates}}\ and\ \bibinfo {author} {\bibfnamefont {W.~H.}\ \bibnamefont
  {Zurek}},\ }\href@noop {} {\bibfield  {journal} {\bibinfo  {journal} {Phys.
  Rev. Lett.}\ }\textbf {\bibinfo {volume} {80}},\ \bibinfo {pages} {5477}
  (\bibinfo {year} {1998})}\BibitemShut {NoStop}%
\bibitem [{\citenamefont {Anglin}\ and\ \citenamefont
  {Zurek}(1999)}]{Anglin99}%
  \BibitemOpen
  \bibfield  {author} {\bibinfo {author} {\bibfnamefont {J.~R.}\ \bibnamefont
  {Anglin}}\ and\ \bibinfo {author} {\bibfnamefont {W.~H.}\ \bibnamefont
  {Zurek}},\ }\href@noop {} {\bibfield  {journal} {\bibinfo  {journal} {Phys.
  Rev. Lett.}\ }\textbf {\bibinfo {volume} {83}},\ \bibinfo {pages} {1707}
  (\bibinfo {year} {1999})}\BibitemShut {NoStop}%
\bibitem [{\citenamefont {Dziarmaga}\ \emph {et~al.}(1999)\citenamefont
  {Dziarmaga}, \citenamefont {Laguna},\ and\ \citenamefont
  {Zurek}}]{Dziarmaga99}%
  \BibitemOpen
  \bibfield  {author} {\bibinfo {author} {\bibfnamefont {J.}~\bibnamefont
  {Dziarmaga}}, \bibinfo {author} {\bibfnamefont {P.}~\bibnamefont {Laguna}}, \
  and\ \bibinfo {author} {\bibfnamefont {W.~H.}\ \bibnamefont {Zurek}},\
  }\href@noop {} {\bibfield  {journal} {\bibinfo  {journal} {Phys. Rev. Lett.}\
  }\textbf {\bibinfo {volume} {82}},\ \bibinfo {pages} {4749} (\bibinfo {year}
  {1999})}\BibitemShut {NoStop}%
\bibitem [{\citenamefont {Antunes}\ \emph {et~al.}(1999)\citenamefont
  {Antunes}, \citenamefont {Bettencourt},\ and\ \citenamefont
  {Zurek}}]{Antunes99}%
  \BibitemOpen
  \bibfield  {author} {\bibinfo {author} {\bibfnamefont {N.~D.}\ \bibnamefont
  {Antunes}}, \bibinfo {author} {\bibfnamefont {L.~M.~A.}\ \bibnamefont
  {Bettencourt}}, \ and\ \bibinfo {author} {\bibfnamefont {W.~H.}\ \bibnamefont
  {Zurek}},\ }\href@noop {} {\bibfield  {journal} {\bibinfo  {journal} {Phys.
  Rev. Lett.}\ }\textbf {\bibinfo {volume} {82}},\ \bibinfo {pages} {2824}
  (\bibinfo {year} {1999})}\BibitemShut {NoStop}%
\bibitem [{\citenamefont {Stephens}\ \emph {et~al.}(1999)\citenamefont
  {Stephens}, \citenamefont {Calzetta}, \citenamefont {Hu},\ and\ \citenamefont
  {Ramsey}}]{Stephens99}%
  \BibitemOpen
  \bibfield  {author} {\bibinfo {author} {\bibfnamefont {G.~J.}\ \bibnamefont
  {Stephens}}, \bibinfo {author} {\bibfnamefont {E.~A.}\ \bibnamefont
  {Calzetta}}, \bibinfo {author} {\bibfnamefont {B.~L.}\ \bibnamefont {Hu}}, \
  and\ \bibinfo {author} {\bibfnamefont {S.~A.}\ \bibnamefont {Ramsey}},\
  }\href@noop {} {\bibfield  {journal} {\bibinfo  {journal} {Phys. Rev. D}\
  }\textbf {\bibinfo {volume} {59}},\ \bibinfo {pages} {045009} (\bibinfo
  {year} {1999})}\BibitemShut {NoStop}%
\bibitem [{\citenamefont {Stephens}\ \emph {et~al.}(2002)\citenamefont
  {Stephens}, \citenamefont {Bettencourt},\ and\ \citenamefont
  {Zurek}}]{Stephens02}%
  \BibitemOpen
  \bibfield  {author} {\bibinfo {author} {\bibfnamefont {G.~J.}\ \bibnamefont
  {Stephens}}, \bibinfo {author} {\bibfnamefont {L.~M.~A.}\ \bibnamefont
  {Bettencourt}}, \ and\ \bibinfo {author} {\bibfnamefont {W.~H.}\ \bibnamefont
  {Zurek}},\ }\href@noop {} {\bibfield  {journal} {\bibinfo  {journal} {Phys.
  Rev. Lett.}\ }\textbf {\bibinfo {volume} {88}},\ \bibinfo {pages} {137004}
  (\bibinfo {year} {2002})}\BibitemShut {NoStop}%
\bibitem [{\citenamefont {Zurek}\ \emph {et~al.}(2005)\citenamefont {Zurek},
  \citenamefont {Dorner},\ and\ \citenamefont {Zoller}}]{Zurek05}%
  \BibitemOpen
  \bibfield  {author} {\bibinfo {author} {\bibfnamefont {W.~H.}\ \bibnamefont
  {Zurek}}, \bibinfo {author} {\bibfnamefont {U.}~\bibnamefont {Dorner}}, \
  and\ \bibinfo {author} {\bibfnamefont {P.}~\bibnamefont {Zoller}},\ }\href
  {\doibase 10.1103/PhysRevLett.95.105701} {\bibfield  {journal} {\bibinfo
  {journal} {Phys. Rev. Lett.}\ }\textbf {\bibinfo {volume} {95}},\ \bibinfo
  {pages} {105701} (\bibinfo {year} {2005})}\BibitemShut {NoStop}%
\bibitem [{\citenamefont {Dziarmaga}(2005)}]{Dziarmaga05}%
  \BibitemOpen
  \bibfield  {author} {\bibinfo {author} {\bibfnamefont {J.}~\bibnamefont
  {Dziarmaga}},\ }\href@noop {} {\bibfield  {journal} {\bibinfo  {journal}
  {Phys. Rev. Lett.}\ }\textbf {\bibinfo {volume} {95}},\ \bibinfo {pages}
  {245701} (\bibinfo {year} {2005})}\BibitemShut {NoStop}%
\bibitem [{\citenamefont {Polkovnikov}(2005)}]{Polkovnikov05}%
  \BibitemOpen
  \bibfield  {author} {\bibinfo {author} {\bibfnamefont {A.}~\bibnamefont
  {Polkovnikov}},\ }\href@noop {} {\bibfield  {journal} {\bibinfo  {journal}
  {Phys. Rev. B}\ }\textbf {\bibinfo {volume} {72}},\ \bibinfo {pages} {161201}
  (\bibinfo {year} {2005})}\BibitemShut {NoStop}%
\bibitem [{\citenamefont {Damski}\ and\ \citenamefont
  {Zurek}(2006)}]{Damski06}%
  \BibitemOpen
  \bibfield  {author} {\bibinfo {author} {\bibfnamefont {B.}~\bibnamefont
  {Damski}}\ and\ \bibinfo {author} {\bibfnamefont {W.~H.}\ \bibnamefont
  {Zurek}},\ }\href@noop {} {\bibfield  {journal} {\bibinfo  {journal} {Phys.
  Rev. A}\ }\textbf {\bibinfo {volume} {73}},\ \bibinfo {pages} {063405}
  (\bibinfo {year} {2006})}\BibitemShut {NoStop}%
\bibitem [{\citenamefont {Cherng}\ and\ \citenamefont
  {Levitov}(2006)}]{Cherng06}%
  \BibitemOpen
  \bibfield  {author} {\bibinfo {author} {\bibfnamefont {R.~W.}\ \bibnamefont
  {Cherng}}\ and\ \bibinfo {author} {\bibfnamefont {L.~S.}\ \bibnamefont
  {Levitov}},\ }\href@noop {} {\bibfield  {journal} {\bibinfo  {journal} {Phys.
  Rev. A}\ }\textbf {\bibinfo {volume} {73}},\ \bibinfo {pages} {043614}
  (\bibinfo {year} {2006})}\BibitemShut {NoStop}%
\bibitem [{\citenamefont {Lamacraft}(2007)}]{Lamacraft07}%
  \BibitemOpen
  \bibfield  {author} {\bibinfo {author} {\bibfnamefont {A.}~\bibnamefont
  {Lamacraft}},\ }\href {\doibase 10.1103/PhysRevLett.98.160404} {\bibfield
  {journal} {\bibinfo  {journal} {Phys. Rev. Lett.}\ }\textbf {\bibinfo
  {volume} {98}},\ \bibinfo {pages} {160404} (\bibinfo {year}
  {2007})}\BibitemShut {NoStop}%
\bibitem [{\citenamefont {Damski}\ and\ \citenamefont
  {Zurek}(2007)}]{Damski07}%
  \BibitemOpen
  \bibfield  {author} {\bibinfo {author} {\bibfnamefont {B.}~\bibnamefont
  {Damski}}\ and\ \bibinfo {author} {\bibfnamefont {W.~H.}\ \bibnamefont
  {Zurek}},\ }\href {\doibase 10.1103/PhysRevLett.99.130402} {\bibfield
  {journal} {\bibinfo  {journal} {Phys. Rev. Lett.}\ }\textbf {\bibinfo
  {volume} {99}},\ \bibinfo {pages} {130402} (\bibinfo {year}
  {2007})}\BibitemShut {NoStop}%
\bibitem [{\citenamefont {Fubini}\ \emph {et~al.}(2007)\citenamefont {Fubini},
  \citenamefont {Falci},\ and\ \citenamefont {Osterloh}}]{Fubini07}%
  \BibitemOpen
  \bibfield  {author} {\bibinfo {author} {\bibfnamefont {A.}~\bibnamefont
  {Fubini}}, \bibinfo {author} {\bibfnamefont {G.}~\bibnamefont {Falci}}, \
  and\ \bibinfo {author} {\bibfnamefont {A.}~\bibnamefont {Osterloh}},\
  }\href@noop {} {\bibfield  {journal} {\bibinfo  {journal} {New J. Phys.}\
  }\textbf {\bibinfo {volume} {9}},\ \bibinfo {pages} {134} (\bibinfo {year}
  {2007})}\BibitemShut {NoStop}%
\bibitem [{\citenamefont {Zurek}(2009)}]{Zurek09}%
  \BibitemOpen
  \bibfield  {author} {\bibinfo {author} {\bibfnamefont {W.~H.}\ \bibnamefont
  {Zurek}},\ }\href@noop {} {\bibfield  {journal} {\bibinfo  {journal} {Phys.
  Rev. Lett.}\ }\textbf {\bibinfo {volume} {102}},\ \bibinfo {pages} {105702}
  (\bibinfo {year} {2009})}\BibitemShut {NoStop}%
\bibitem [{\citenamefont {Dziarmaga}(2010)}]{Dziarmaga10}%
  \BibitemOpen
  \bibfield  {author} {\bibinfo {author} {\bibfnamefont {J.}~\bibnamefont
  {Dziarmaga}},\ }\href@noop {} {\bibfield  {journal} {\bibinfo  {journal}
  {Adv. Phys.}\ }\textbf {\bibinfo {volume} {59}},\ \bibinfo {pages} {1063}
  (\bibinfo {year} {2010})}\BibitemShut {NoStop}%
\bibitem [{\citenamefont {Chuang}\ \emph {et~al.}(1991)\citenamefont {Chuang},
  \citenamefont {Durrer}, \citenamefont {Turok},\ and\ \citenamefont
  {Yurke}}]{Chuang91}%
  \BibitemOpen
  \bibfield  {author} {\bibinfo {author} {\bibfnamefont {I.}~\bibnamefont
  {Chuang}}, \bibinfo {author} {\bibfnamefont {R.}~\bibnamefont {Durrer}},
  \bibinfo {author} {\bibfnamefont {N.}~\bibnamefont {Turok}}, \ and\ \bibinfo
  {author} {\bibfnamefont {B.}~\bibnamefont {Yurke}},\ }\href {\doibase
  10.1126/science.251.4999.1336} {\bibfield  {journal} {\bibinfo  {journal}
  {Science}\ }\textbf {\bibinfo {volume} {251}},\ \bibinfo {pages} {1336}
  (\bibinfo {year} {1991})}\BibitemShut {NoStop}%
\bibitem [{\citenamefont {Bowick}\ \emph {et~al.}(1994)\citenamefont {Bowick},
  \citenamefont {Chandar}, \citenamefont {Schiff},\ and\ \citenamefont
  {Srivastava}}]{Bowick94}%
  \BibitemOpen
  \bibfield  {author} {\bibinfo {author} {\bibfnamefont {M.~J.}\ \bibnamefont
  {Bowick}}, \bibinfo {author} {\bibfnamefont {L.}~\bibnamefont {Chandar}},
  \bibinfo {author} {\bibfnamefont {E.~A.}\ \bibnamefont {Schiff}}, \ and\
  \bibinfo {author} {\bibfnamefont {A.~M.}\ \bibnamefont {Srivastava}},\ }\href
  {\doibase 10.1126/science.263.5149.943} {\bibfield  {journal} {\bibinfo
  {journal} {Science}\ }\textbf {\bibinfo {volume} {263}},\ \bibinfo {pages}
  {943} (\bibinfo {year} {1994})}\BibitemShut {NoStop}%
\bibitem [{\citenamefont {Hendry}\ \emph {et~al.}(1994)\citenamefont {Hendry},
  \citenamefont {Lawson}, \citenamefont {Lee}, \citenamefont {McClintock},\
  and\ \citenamefont {Williams}}]{Hendry94}%
  \BibitemOpen
  \bibfield  {author} {\bibinfo {author} {\bibfnamefont {P.}~\bibnamefont
  {Hendry}}, \bibinfo {author} {\bibfnamefont {N.}~\bibnamefont {Lawson}},
  \bibinfo {author} {\bibfnamefont {R.}~\bibnamefont {Lee}}, \bibinfo {author}
  {\bibfnamefont {P.~V.}\ \bibnamefont {McClintock}}, \ and\ \bibinfo {author}
  {\bibfnamefont {C.}~\bibnamefont {Williams}},\ }\href@noop {} {\bibfield
  {journal} {\bibinfo  {journal} {Nature}\ }\textbf {\bibinfo {volume} {368}},\
  \bibinfo {pages} {315} (\bibinfo {year} {1994})}\BibitemShut {NoStop}%
\bibitem [{\citenamefont {Ruutu}\ \emph {et~al.}(1996)\citenamefont {Ruutu},
  \citenamefont {Eltsov}, \citenamefont {Gill}, \citenamefont {Kibble},
  \citenamefont {Krusius}, \citenamefont {Makhlin}, \citenamefont {Placais},
  \citenamefont {Volovik},\ and\ \citenamefont {Xu}}]{Ruutu96}%
  \BibitemOpen
  \bibfield  {author} {\bibinfo {author} {\bibfnamefont {V.}~\bibnamefont
  {Ruutu}}, \bibinfo {author} {\bibfnamefont {V.}~\bibnamefont {Eltsov}},
  \bibinfo {author} {\bibfnamefont {A.}~\bibnamefont {Gill}}, \bibinfo {author}
  {\bibfnamefont {T.}~\bibnamefont {Kibble}}, \bibinfo {author} {\bibfnamefont
  {M.}~\bibnamefont {Krusius}}, \bibinfo {author} {\bibfnamefont {Y.~G.}\
  \bibnamefont {Makhlin}}, \bibinfo {author} {\bibfnamefont {B.}~\bibnamefont
  {Placais}}, \bibinfo {author} {\bibfnamefont {G.}~\bibnamefont {Volovik}}, \
  and\ \bibinfo {author} {\bibfnamefont {W.}~\bibnamefont {Xu}},\ }\href@noop
  {} {\bibfield  {journal} {\bibinfo  {journal} {Nature}\ }\textbf {\bibinfo
  {volume} {382}},\ \bibinfo {pages} {334} (\bibinfo {year}
  {1996})}\BibitemShut {NoStop}%
\bibitem [{\citenamefont {B{\"a}uerle}\ \emph {et~al.}(1996)\citenamefont
  {B{\"a}uerle}, \citenamefont {Bunkov}, \citenamefont {Fisher}, \citenamefont
  {Godfrin},\ and\ \citenamefont {Pickett}}]{Bauerle96}%
  \BibitemOpen
  \bibfield  {author} {\bibinfo {author} {\bibfnamefont {C.}~\bibnamefont
  {B{\"a}uerle}}, \bibinfo {author} {\bibfnamefont {Y.~M.}\ \bibnamefont
  {Bunkov}}, \bibinfo {author} {\bibfnamefont {S.}~\bibnamefont {Fisher}},
  \bibinfo {author} {\bibfnamefont {H.}~\bibnamefont {Godfrin}}, \ and\
  \bibinfo {author} {\bibfnamefont {G.}~\bibnamefont {Pickett}},\ }\href@noop
  {} {\bibfield  {journal} {\bibinfo  {journal} {Nature}\ }\textbf {\bibinfo
  {volume} {382}},\ \bibinfo {pages} {332} (\bibinfo {year}
  {1996})}\BibitemShut {NoStop}%
\bibitem [{\citenamefont {Ducci}\ \emph {et~al.}(1999)\citenamefont {Ducci},
  \citenamefont {Ramazza}, \citenamefont {Gonz\'alez-Vi\~nas},\ and\
  \citenamefont {Arecchi}}]{Ducci99}%
  \BibitemOpen
  \bibfield  {author} {\bibinfo {author} {\bibfnamefont {S.}~\bibnamefont
  {Ducci}}, \bibinfo {author} {\bibfnamefont {P.~L.}\ \bibnamefont {Ramazza}},
  \bibinfo {author} {\bibfnamefont {W.}~\bibnamefont {Gonz\'alez-Vi\~nas}}, \
  and\ \bibinfo {author} {\bibfnamefont {F.~T.}\ \bibnamefont {Arecchi}},\
  }\href {\doibase 10.1103/PhysRevLett.83.5210} {\bibfield  {journal} {\bibinfo
   {journal} {Phys. Rev. Lett.}\ }\textbf {\bibinfo {volume} {83}},\ \bibinfo
  {pages} {5210} (\bibinfo {year} {1999})}\BibitemShut {NoStop}%
\bibitem [{\citenamefont {Carmi}\ \emph {et~al.}(2000)\citenamefont {Carmi},
  \citenamefont {Polturak},\ and\ \citenamefont {Koren}}]{Carmi00}%
  \BibitemOpen
  \bibfield  {author} {\bibinfo {author} {\bibfnamefont {R.}~\bibnamefont
  {Carmi}}, \bibinfo {author} {\bibfnamefont {E.}~\bibnamefont {Polturak}}, \
  and\ \bibinfo {author} {\bibfnamefont {G.}~\bibnamefont {Koren}},\ }\href
  {\doibase 10.1103/PhysRevLett.84.4966} {\bibfield  {journal} {\bibinfo
  {journal} {Phys. Rev. Lett.}\ }\textbf {\bibinfo {volume} {84}},\ \bibinfo
  {pages} {4966} (\bibinfo {year} {2000})}\BibitemShut {NoStop}%
\bibitem [{\citenamefont {Maniv}\ \emph {et~al.}(2003)\citenamefont {Maniv},
  \citenamefont {Polturak},\ and\ \citenamefont {Koren}}]{Maniv03}%
  \BibitemOpen
  \bibfield  {author} {\bibinfo {author} {\bibfnamefont {A.}~\bibnamefont
  {Maniv}}, \bibinfo {author} {\bibfnamefont {E.}~\bibnamefont {Polturak}}, \
  and\ \bibinfo {author} {\bibfnamefont {G.}~\bibnamefont {Koren}},\ }\href
  {\doibase 10.1103/PhysRevLett.91.197001} {\bibfield  {journal} {\bibinfo
  {journal} {Phys. Rev. Lett.}\ }\textbf {\bibinfo {volume} {91}},\ \bibinfo
  {pages} {197001} (\bibinfo {year} {2003})}\BibitemShut {NoStop}%
\bibitem [{\citenamefont {Monaco}\ \emph {et~al.}(2006)\citenamefont {Monaco},
  \citenamefont {Chumakov}, \citenamefont {Yue}, \citenamefont {Monaco},
  \citenamefont {Comez}, \citenamefont {Fioretto}, \citenamefont {Crichton},\
  and\ \citenamefont {R\"uffer}}]{Monaco06}%
  \BibitemOpen
  \bibfield  {author} {\bibinfo {author} {\bibfnamefont {A.}~\bibnamefont
  {Monaco}}, \bibinfo {author} {\bibfnamefont {A.~I.}\ \bibnamefont
  {Chumakov}}, \bibinfo {author} {\bibfnamefont {Y.-Z.}\ \bibnamefont {Yue}},
  \bibinfo {author} {\bibfnamefont {G.}~\bibnamefont {Monaco}}, \bibinfo
  {author} {\bibfnamefont {L.}~\bibnamefont {Comez}}, \bibinfo {author}
  {\bibfnamefont {D.}~\bibnamefont {Fioretto}}, \bibinfo {author}
  {\bibfnamefont {W.~A.}\ \bibnamefont {Crichton}}, \ and\ \bibinfo {author}
  {\bibfnamefont {R.}~\bibnamefont {R\"uffer}},\ }\href {\doibase
  10.1103/PhysRevLett.96.205502} {\bibfield  {journal} {\bibinfo  {journal}
  {Phys. Rev. Lett.}\ }\textbf {\bibinfo {volume} {96}},\ \bibinfo {pages}
  {205502} (\bibinfo {year} {2006})}\BibitemShut {NoStop}%
\bibitem [{\citenamefont {del Campo}\ \emph {et~al.}(2010)\citenamefont {del
  Campo}, \citenamefont {De~Chiara}, \citenamefont {Morigi}, \citenamefont
  {Plenio},\ and\ \citenamefont {Retzker}}]{delCampo10}%
  \BibitemOpen
  \bibfield  {author} {\bibinfo {author} {\bibfnamefont {A.}~\bibnamefont {del
  Campo}}, \bibinfo {author} {\bibfnamefont {G.}~\bibnamefont {De~Chiara}},
  \bibinfo {author} {\bibfnamefont {G.}~\bibnamefont {Morigi}}, \bibinfo
  {author} {\bibfnamefont {M.~B.}\ \bibnamefont {Plenio}}, \ and\ \bibinfo
  {author} {\bibfnamefont {A.}~\bibnamefont {Retzker}},\ }\href@noop {}
  {\bibfield  {journal} {\bibinfo  {journal} {Phys. Rev. Lett.}\ }\textbf
  {\bibinfo {volume} {105}},\ \bibinfo {pages} {075701} (\bibinfo {year}
  {2010})}\BibitemShut {NoStop}%
\bibitem [{\citenamefont {Pyka}\ \emph {et~al.}(2013)\citenamefont {Pyka},
  \citenamefont {Keller}, \citenamefont {Partner}, \citenamefont {Nigmatullin},
  \citenamefont {Burgermeister}, \citenamefont {Meier}, \citenamefont
  {Kuhlmann}, \citenamefont {Retzker}, \citenamefont {Plenio}, \citenamefont
  {Zurek} \emph {et~al.}}]{Pyka13}%
  \BibitemOpen
  \bibfield  {author} {\bibinfo {author} {\bibfnamefont {K.}~\bibnamefont
  {Pyka}}, \bibinfo {author} {\bibfnamefont {J.}~\bibnamefont {Keller}},
  \bibinfo {author} {\bibfnamefont {H.}~\bibnamefont {Partner}}, \bibinfo
  {author} {\bibfnamefont {R.}~\bibnamefont {Nigmatullin}}, \bibinfo {author}
  {\bibfnamefont {T.}~\bibnamefont {Burgermeister}}, \bibinfo {author}
  {\bibfnamefont {D.}~\bibnamefont {Meier}}, \bibinfo {author} {\bibfnamefont
  {K.}~\bibnamefont {Kuhlmann}}, \bibinfo {author} {\bibfnamefont
  {A.}~\bibnamefont {Retzker}}, \bibinfo {author} {\bibfnamefont
  {M.}~\bibnamefont {Plenio}}, \bibinfo {author} {\bibfnamefont
  {W.}~\bibnamefont {Zurek}},  \emph {et~al.},\ }\href@noop {} {\bibfield
  {journal} {\bibinfo  {journal} {Nat. Commun.}\ }\textbf {\bibinfo {volume}
  {4}} (\bibinfo {year} {2013})}\BibitemShut {NoStop}%
\bibitem [{\citenamefont {Ejtemaee}\ and\ \citenamefont
  {Haljan}(2013)}]{Ejtemaee13}%
  \BibitemOpen
  \bibfield  {author} {\bibinfo {author} {\bibfnamefont {S.}~\bibnamefont
  {Ejtemaee}}\ and\ \bibinfo {author} {\bibfnamefont {P.~C.}\ \bibnamefont
  {Haljan}},\ }\href@noop {} {\bibfield  {journal} {\bibinfo  {journal} {Phys.
  Rev. A}\ }\textbf {\bibinfo {volume} {87}},\ \bibinfo {pages} {051401}
  (\bibinfo {year} {2013})}\BibitemShut {NoStop}%
\bibitem [{\citenamefont {Ulm}\ \emph {et~al.}(2013)\citenamefont {Ulm},
  \citenamefont {Ro{\ss}nagel}, \citenamefont {Jacob}, \citenamefont
  {Deg{\"u}nther}, \citenamefont {Dawkins}, \citenamefont {Poschinger},
  \citenamefont {Nigmatullin}, \citenamefont {Retzker}, \citenamefont {Plenio},
  \citenamefont {Schmidt-Kaler} \emph {et~al.}}]{Ulm13}%
  \BibitemOpen
  \bibfield  {author} {\bibinfo {author} {\bibfnamefont {S.}~\bibnamefont
  {Ulm}}, \bibinfo {author} {\bibfnamefont {J.}~\bibnamefont {Ro{\ss}nagel}},
  \bibinfo {author} {\bibfnamefont {G.}~\bibnamefont {Jacob}}, \bibinfo
  {author} {\bibfnamefont {C.}~\bibnamefont {Deg{\"u}nther}}, \bibinfo {author}
  {\bibfnamefont {S.}~\bibnamefont {Dawkins}}, \bibinfo {author} {\bibfnamefont
  {U.}~\bibnamefont {Poschinger}}, \bibinfo {author} {\bibfnamefont
  {R.}~\bibnamefont {Nigmatullin}}, \bibinfo {author} {\bibfnamefont
  {A.}~\bibnamefont {Retzker}}, \bibinfo {author} {\bibfnamefont
  {M.}~\bibnamefont {Plenio}}, \bibinfo {author} {\bibfnamefont
  {F.}~\bibnamefont {Schmidt-Kaler}},  \emph {et~al.},\ }\href@noop {}
  {\bibfield  {journal} {\bibinfo  {journal} {Nat. Commun.}\ }\textbf {\bibinfo
  {volume} {4}} (\bibinfo {year} {2013})}\BibitemShut {NoStop}%
\bibitem [{\citenamefont {Chen}\ \emph {et~al.}(2011)\citenamefont {Chen},
  \citenamefont {White}, \citenamefont {Borries},\ and\ \citenamefont
  {DeMarco}}]{Chen11}%
  \BibitemOpen
  \bibfield  {author} {\bibinfo {author} {\bibfnamefont {D.}~\bibnamefont
  {Chen}}, \bibinfo {author} {\bibfnamefont {M.}~\bibnamefont {White}},
  \bibinfo {author} {\bibfnamefont {C.}~\bibnamefont {Borries}}, \ and\
  \bibinfo {author} {\bibfnamefont {B.}~\bibnamefont {DeMarco}},\ }\href
  {\doibase 10.1103/PhysRevLett.106.235304} {\bibfield  {journal} {\bibinfo
  {journal} {Phys. Rev. Lett.}\ }\textbf {\bibinfo {volume} {106}},\ \bibinfo
  {pages} {235304} (\bibinfo {year} {2011})}\BibitemShut {NoStop}%
\bibitem [{\citenamefont {Donner}\ \emph {et~al.}(2007)\citenamefont {Donner},
  \citenamefont {Ritter}, \citenamefont {Bourdel}, \citenamefont {{\"O}ttl},
  \citenamefont {K{\"o}hl},\ and\ \citenamefont {Esslinger}}]{Donner07}%
  \BibitemOpen
  \bibfield  {author} {\bibinfo {author} {\bibfnamefont {T.}~\bibnamefont
  {Donner}}, \bibinfo {author} {\bibfnamefont {S.}~\bibnamefont {Ritter}},
  \bibinfo {author} {\bibfnamefont {T.}~\bibnamefont {Bourdel}}, \bibinfo
  {author} {\bibfnamefont {A.}~\bibnamefont {{\"O}ttl}}, \bibinfo {author}
  {\bibfnamefont {M.}~\bibnamefont {K{\"o}hl}}, \ and\ \bibinfo {author}
  {\bibfnamefont {T.}~\bibnamefont {Esslinger}},\ }\href@noop {} {\bibfield
  {journal} {\bibinfo  {journal} {Science}\ }\textbf {\bibinfo {volume}
  {315}},\ \bibinfo {pages} {1556} (\bibinfo {year} {2007})}\BibitemShut
  {NoStop}%
\bibitem [{\citenamefont {Navon}\ \emph {et~al.}(2015)\citenamefont {Navon},
  \citenamefont {Gaunt}, \citenamefont {Smith},\ and\ \citenamefont
  {Hadzibabic}}]{Navon15}%
  \BibitemOpen
  \bibfield  {author} {\bibinfo {author} {\bibfnamefont {N.}~\bibnamefont
  {Navon}}, \bibinfo {author} {\bibfnamefont {A.~L.}\ \bibnamefont {Gaunt}},
  \bibinfo {author} {\bibfnamefont {R.~P.}\ \bibnamefont {Smith}}, \ and\
  \bibinfo {author} {\bibfnamefont {Z.}~\bibnamefont {Hadzibabic}},\
  }\href@noop {} {\bibfield  {journal} {\bibinfo  {journal} {Science}\ }\textbf
  {\bibinfo {volume} {347}},\ \bibinfo {pages} {167} (\bibinfo {year}
  {2015})}\BibitemShut {NoStop}%
\bibitem [{\citenamefont {Scherer}\ \emph {et~al.}(2007)\citenamefont
  {Scherer}, \citenamefont {Weiler}, \citenamefont {Neely},\ and\ \citenamefont
  {Anderson}}]{Scherer07}%
  \BibitemOpen
  \bibfield  {author} {\bibinfo {author} {\bibfnamefont {D.~R.}\ \bibnamefont
  {Scherer}}, \bibinfo {author} {\bibfnamefont {C.~N.}\ \bibnamefont {Weiler}},
  \bibinfo {author} {\bibfnamefont {T.~W.}\ \bibnamefont {Neely}}, \ and\
  \bibinfo {author} {\bibfnamefont {B.~P.}\ \bibnamefont {Anderson}},\ }\href
  {\doibase 10.1103/PhysRevLett.98.110402} {\bibfield  {journal} {\bibinfo
  {journal} {Phys. Rev. Lett.}\ }\textbf {\bibinfo {volume} {98}},\ \bibinfo
  {pages} {110402} (\bibinfo {year} {2007})}\BibitemShut {NoStop}%
\bibitem [{\citenamefont {Weiler}\ \emph {et~al.}(2008)\citenamefont {Weiler},
  \citenamefont {Neely}, \citenamefont {Scherer}, \citenamefont {Bradley},
  \citenamefont {Davis},\ and\ \citenamefont {Anderson}}]{Weiler08}%
  \BibitemOpen
  \bibfield  {author} {\bibinfo {author} {\bibfnamefont {C.~N.}\ \bibnamefont
  {Weiler}}, \bibinfo {author} {\bibfnamefont {T.~W.}\ \bibnamefont {Neely}},
  \bibinfo {author} {\bibfnamefont {D.~R.}\ \bibnamefont {Scherer}}, \bibinfo
  {author} {\bibfnamefont {A.~S.}\ \bibnamefont {Bradley}}, \bibinfo {author}
  {\bibfnamefont {M.~J.}\ \bibnamefont {Davis}}, \ and\ \bibinfo {author}
  {\bibfnamefont {B.~P.}\ \bibnamefont {Anderson}},\ }\href@noop {} {\bibfield
  {journal} {\bibinfo  {journal} {Nature}\ }\textbf {\bibinfo {volume} {455}},\
  \bibinfo {pages} {948} (\bibinfo {year} {2008})}\BibitemShut {NoStop}%
\bibitem [{\citenamefont {Lamporesi}\ \emph {et~al.}(2013)\citenamefont
  {Lamporesi}, \citenamefont {Donadello}, \citenamefont {Serafini},
  \citenamefont {Dalfovo},\ and\ \citenamefont {Ferrari}}]{Lamporesi13}%
  \BibitemOpen
  \bibfield  {author} {\bibinfo {author} {\bibfnamefont {G.}~\bibnamefont
  {Lamporesi}}, \bibinfo {author} {\bibfnamefont {S.}~\bibnamefont
  {Donadello}}, \bibinfo {author} {\bibfnamefont {S.}~\bibnamefont {Serafini}},
  \bibinfo {author} {\bibfnamefont {F.}~\bibnamefont {Dalfovo}}, \ and\
  \bibinfo {author} {\bibfnamefont {G.}~\bibnamefont {Ferrari}},\ }\href@noop
  {} {\bibfield  {journal} {\bibinfo  {journal} {Nature Phys.}\ }\textbf
  {\bibinfo {volume} {9}},\ \bibinfo {pages} {656} (\bibinfo {year}
  {2013})}\BibitemShut {NoStop}%
\bibitem [{\citenamefont {Corman}\ \emph {et~al.}(2014)\citenamefont {Corman},
  \citenamefont {Chomaz}, \citenamefont {Bienaim{\'e}}, \citenamefont
  {Desbuquois}, \citenamefont {Weitenberg}, \citenamefont {Nascimbene},
  \citenamefont {Dalibard},\ and\ \citenamefont {Beugnon}}]{Corman14}%
  \BibitemOpen
  \bibfield  {author} {\bibinfo {author} {\bibfnamefont {L.}~\bibnamefont
  {Corman}}, \bibinfo {author} {\bibfnamefont {L.}~\bibnamefont {Chomaz}},
  \bibinfo {author} {\bibfnamefont {T.}~\bibnamefont {Bienaim{\'e}}}, \bibinfo
  {author} {\bibfnamefont {R.}~\bibnamefont {Desbuquois}}, \bibinfo {author}
  {\bibfnamefont {C.}~\bibnamefont {Weitenberg}}, \bibinfo {author}
  {\bibfnamefont {S.}~\bibnamefont {Nascimbene}}, \bibinfo {author}
  {\bibfnamefont {J.}~\bibnamefont {Dalibard}}, \ and\ \bibinfo {author}
  {\bibfnamefont {J.}~\bibnamefont {Beugnon}},\ }\href@noop {} {\bibfield
  {journal} {\bibinfo  {journal} {Phys. Rev. Lett.}\ }\textbf {\bibinfo
  {volume} {113}},\ \bibinfo {pages} {135302} (\bibinfo {year}
  {2014})}\BibitemShut {NoStop}%
\bibitem [{\citenamefont {Braun}\ \emph {et~al.}(2015)\citenamefont {Braun},
  \citenamefont {Friesdorf}, \citenamefont {Hodgman}, \citenamefont
  {Schreiber}, \citenamefont {Ronzheimer}, \citenamefont {Riera}, \citenamefont
  {del Rey}, \citenamefont {Bloch}, \citenamefont {Eisert},\ and\ \citenamefont
  {Schneider}}]{Braun15}%
  \BibitemOpen
  \bibfield  {author} {\bibinfo {author} {\bibfnamefont {S.}~\bibnamefont
  {Braun}}, \bibinfo {author} {\bibfnamefont {M.}~\bibnamefont {Friesdorf}},
  \bibinfo {author} {\bibfnamefont {S.~S.}\ \bibnamefont {Hodgman}}, \bibinfo
  {author} {\bibfnamefont {M.}~\bibnamefont {Schreiber}}, \bibinfo {author}
  {\bibfnamefont {J.~P.}\ \bibnamefont {Ronzheimer}}, \bibinfo {author}
  {\bibfnamefont {A.}~\bibnamefont {Riera}}, \bibinfo {author} {\bibfnamefont
  {M.}~\bibnamefont {del Rey}}, \bibinfo {author} {\bibfnamefont
  {I.}~\bibnamefont {Bloch}}, \bibinfo {author} {\bibfnamefont
  {J.}~\bibnamefont {Eisert}}, \ and\ \bibinfo {author} {\bibfnamefont
  {U.}~\bibnamefont {Schneider}},\ }\href@noop {} {\bibfield  {journal}
  {\bibinfo  {journal} {Proceedings of the National Academy of Sciences}\
  }\textbf {\bibinfo {volume} {112}},\ \bibinfo {pages} {3641} (\bibinfo {year}
  {2015})}\BibitemShut {NoStop}%
\bibitem [{\citenamefont {Cui}\ \emph {et~al.}(2015)\citenamefont {Cui},
  \citenamefont {Huang}, \citenamefont {Wang}, \citenamefont {Cao},
  \citenamefont {Wang}, \citenamefont {Lv}, \citenamefont {Lu}, \citenamefont
  {Luo}, \citenamefont {del Campo}, \citenamefont {Han} \emph
  {et~al.}}]{Cui15}%
  \BibitemOpen
  \bibfield  {author} {\bibinfo {author} {\bibfnamefont {J.-M.}\ \bibnamefont
  {Cui}}, \bibinfo {author} {\bibfnamefont {Y.-F.}\ \bibnamefont {Huang}},
  \bibinfo {author} {\bibfnamefont {Z.}~\bibnamefont {Wang}}, \bibinfo {author}
  {\bibfnamefont {D.-Y.}\ \bibnamefont {Cao}}, \bibinfo {author} {\bibfnamefont
  {J.}~\bibnamefont {Wang}}, \bibinfo {author} {\bibfnamefont {W.-M.}\
  \bibnamefont {Lv}}, \bibinfo {author} {\bibfnamefont {Y.}~\bibnamefont {Lu}},
  \bibinfo {author} {\bibfnamefont {L.}~\bibnamefont {Luo}}, \bibinfo {author}
  {\bibfnamefont {A.}~\bibnamefont {del Campo}}, \bibinfo {author}
  {\bibfnamefont {Y.-J.}\ \bibnamefont {Han}},  \emph {et~al.},\ }\href@noop {}
  {\enquote {\bibinfo {title} {Supporting kibble-zurek mechanism in quantum
  ising model through a trapped ion},}\ } (\bibinfo {year} {2015}),\ \bibinfo
  {note} {arXiv:1505.05734}\BibitemShut {NoStop}%
\bibitem [{\citenamefont {Zhang}\ \emph {et~al.}(2012)\citenamefont {Zhang},
  \citenamefont {Hung}, \citenamefont {Tung},\ and\ \citenamefont
  {Chin}}]{Zhang12}%
  \BibitemOpen
  \bibfield  {author} {\bibinfo {author} {\bibfnamefont {X.}~\bibnamefont
  {Zhang}}, \bibinfo {author} {\bibfnamefont {C.-L.}\ \bibnamefont {Hung}},
  \bibinfo {author} {\bibfnamefont {S.-K.}\ \bibnamefont {Tung}}, \ and\
  \bibinfo {author} {\bibfnamefont {C.}~\bibnamefont {Chin}},\ }\href@noop {}
  {\bibfield  {journal} {\bibinfo  {journal} {Science}\ }\textbf {\bibinfo
  {volume} {335}},\ \bibinfo {pages} {1070} (\bibinfo {year}
  {2012})}\BibitemShut {NoStop}%
\bibitem [{\citenamefont {Nicklas}\ \emph {et~al.}(2015)\citenamefont
  {Nicklas}, \citenamefont {Karl}, \citenamefont {H\"ofer}, \citenamefont
  {Johnson}, \citenamefont {Muessel}, \citenamefont {Strobel}, \citenamefont
  {Tomkovi\ifmmode~\check{c}\else \v{c}\fi{}}, \citenamefont {Gasenzer},\ and\
  \citenamefont {Oberthaler}}]{Nicklas15}%
  \BibitemOpen
  \bibfield  {author} {\bibinfo {author} {\bibfnamefont {E.}~\bibnamefont
  {Nicklas}}, \bibinfo {author} {\bibfnamefont {M.}~\bibnamefont {Karl}},
  \bibinfo {author} {\bibfnamefont {M.}~\bibnamefont {H\"ofer}}, \bibinfo
  {author} {\bibfnamefont {A.}~\bibnamefont {Johnson}}, \bibinfo {author}
  {\bibfnamefont {W.}~\bibnamefont {Muessel}}, \bibinfo {author} {\bibfnamefont
  {H.}~\bibnamefont {Strobel}}, \bibinfo {author} {\bibfnamefont
  {J.}~\bibnamefont {Tomkovi\ifmmode~\check{c}\else \v{c}\fi{}}}, \bibinfo
  {author} {\bibfnamefont {T.}~\bibnamefont {Gasenzer}}, \ and\ \bibinfo
  {author} {\bibfnamefont {M.~K.}\ \bibnamefont {Oberthaler}},\ }\href
  {\doibase 10.1103/PhysRevLett.115.245301} {\bibfield  {journal} {\bibinfo
  {journal} {Phys. Rev. Lett.}\ }\textbf {\bibinfo {volume} {115}},\ \bibinfo
  {pages} {245301} (\bibinfo {year} {2015})}\BibitemShut {NoStop}%
\bibitem [{\citenamefont {Murata}\ \emph {et~al.}(2007)\citenamefont {Murata},
  \citenamefont {Saito},\ and\ \citenamefont {Ueda}}]{Murata07}%
  \BibitemOpen
  \bibfield  {author} {\bibinfo {author} {\bibfnamefont {K.}~\bibnamefont
  {Murata}}, \bibinfo {author} {\bibfnamefont {H.}~\bibnamefont {Saito}}, \
  and\ \bibinfo {author} {\bibfnamefont {M.}~\bibnamefont {Ueda}},\ }\href
  {\doibase 10.1103/PhysRevA.75.013607} {\bibfield  {journal} {\bibinfo
  {journal} {Phys. Rev. A}\ }\textbf {\bibinfo {volume} {75}},\ \bibinfo
  {pages} {013607} (\bibinfo {year} {2007})}\BibitemShut {NoStop}%
\bibitem [{\citenamefont {Damski}\ and\ \citenamefont
  {Zurek}(2008)}]{Damski08}%
  \BibitemOpen
  \bibfield  {author} {\bibinfo {author} {\bibfnamefont {B.}~\bibnamefont
  {Damski}}\ and\ \bibinfo {author} {\bibfnamefont {W.~H.}\ \bibnamefont
  {Zurek}},\ }\href@noop {} {\bibfield  {journal} {\bibinfo  {journal} {New J.
  Phys.}\ }\textbf {\bibinfo {volume} {10}},\ \bibinfo {pages} {045023}
  (\bibinfo {year} {2008})}\BibitemShut {NoStop}%
\bibitem [{\citenamefont {Saito}\ \emph {et~al.}(2007)\citenamefont {Saito},
  \citenamefont {Kawaguchi},\ and\ \citenamefont {Ueda}}]{Saito07}%
  \BibitemOpen
  \bibfield  {author} {\bibinfo {author} {\bibfnamefont {H.}~\bibnamefont
  {Saito}}, \bibinfo {author} {\bibfnamefont {Y.}~\bibnamefont {Kawaguchi}}, \
  and\ \bibinfo {author} {\bibfnamefont {M.}~\bibnamefont {Ueda}},\ }\href
  {\doibase 10.1103/PhysRevA.76.043613} {\bibfield  {journal} {\bibinfo
  {journal} {Phys. Rev. A}\ }\textbf {\bibinfo {volume} {76}},\ \bibinfo
  {pages} {043613} (\bibinfo {year} {2007})}\BibitemShut {NoStop}%
\bibitem [{\citenamefont {Saito}\ \emph {et~al.}(2013)\citenamefont {Saito},
  \citenamefont {Kawaguchi},\ and\ \citenamefont {Ueda}}]{Saito13}%
  \BibitemOpen
  \bibfield  {author} {\bibinfo {author} {\bibfnamefont {H.}~\bibnamefont
  {Saito}}, \bibinfo {author} {\bibfnamefont {Y.}~\bibnamefont {Kawaguchi}}, \
  and\ \bibinfo {author} {\bibfnamefont {M.}~\bibnamefont {Ueda}},\ }\href@noop
  {} {\bibfield  {journal} {\bibinfo  {journal} {J. Phys. Condens. Matter}\
  }\textbf {\bibinfo {volume} {25}},\ \bibinfo {pages} {404212} (\bibinfo
  {year} {2013})}\BibitemShut {NoStop}%
\bibitem [{\citenamefont {Damski}\ and\ \citenamefont
  {Zurek}(2009)}]{Damski09}%
  \BibitemOpen
  \bibfield  {author} {\bibinfo {author} {\bibfnamefont {B.}~\bibnamefont
  {Damski}}\ and\ \bibinfo {author} {\bibfnamefont {W.~H.}\ \bibnamefont
  {Zurek}},\ }\href@noop {} {\bibfield  {journal} {\bibinfo  {journal} {New J.
  Phys.}\ }\textbf {\bibinfo {volume} {11}},\ \bibinfo {pages} {063014}
  (\bibinfo {year} {2009})}\BibitemShut {NoStop}%
\bibitem [{\citenamefont {Sadler}\ \emph {et~al.}(2006)\citenamefont {Sadler},
  \citenamefont {Higbie}, \citenamefont {Leslie}, \citenamefont
  {Vengalattore},\ and\ \citenamefont {Stamper-Kurn}}]{Sadler06}%
  \BibitemOpen
  \bibfield  {author} {\bibinfo {author} {\bibfnamefont {L.~E.}\ \bibnamefont
  {Sadler}}, \bibinfo {author} {\bibfnamefont {J.~M.}\ \bibnamefont {Higbie}},
  \bibinfo {author} {\bibfnamefont {S.~R.}\ \bibnamefont {Leslie}}, \bibinfo
  {author} {\bibfnamefont {M.}~\bibnamefont {Vengalattore}}, \ and\ \bibinfo
  {author} {\bibfnamefont {D.~M.}\ \bibnamefont {Stamper-Kurn}},\ }\href@noop
  {} {\bibfield  {journal} {\bibinfo  {journal} {Nature}\ }\textbf {\bibinfo
  {volume} {443}},\ \bibinfo {pages} {312} (\bibinfo {year}
  {2006})}\BibitemShut {NoStop}%
\bibitem [{\citenamefont {Hamley}\ \emph {et~al.}(2012)\citenamefont {Hamley},
  \citenamefont {Gerving}, \citenamefont {Hoang}, \citenamefont {Bookjans},\
  and\ \citenamefont {Chapman}}]{Hamley12}%
  \BibitemOpen
  \bibfield  {author} {\bibinfo {author} {\bibfnamefont {C.}~\bibnamefont
  {Hamley}}, \bibinfo {author} {\bibfnamefont {C.}~\bibnamefont {Gerving}},
  \bibinfo {author} {\bibfnamefont {T.}~\bibnamefont {Hoang}}, \bibinfo
  {author} {\bibfnamefont {E.}~\bibnamefont {Bookjans}}, \ and\ \bibinfo
  {author} {\bibfnamefont {M.}~\bibnamefont {Chapman}},\ }\href@noop {}
  {\bibfield  {journal} {\bibinfo  {journal} {Nature Phys.}\ }\textbf {\bibinfo
  {volume} {8}},\ \bibinfo {pages} {305} (\bibinfo {year} {2012})}\BibitemShut
  {NoStop}%
\bibitem [{\citenamefont {Gerving}\ \emph {et~al.}(2012)\citenamefont
  {Gerving}, \citenamefont {Hoang}, \citenamefont {Land}, \citenamefont
  {Anquez}, \citenamefont {Hamley},\ and\ \citenamefont {Chapman}}]{Gerving12}%
  \BibitemOpen
  \bibfield  {author} {\bibinfo {author} {\bibfnamefont {C.~S.}\ \bibnamefont
  {Gerving}}, \bibinfo {author} {\bibfnamefont {T.~M.}\ \bibnamefont {Hoang}},
  \bibinfo {author} {\bibfnamefont {B.~J.}\ \bibnamefont {Land}}, \bibinfo
  {author} {\bibfnamefont {M.}~\bibnamefont {Anquez}}, \bibinfo {author}
  {\bibfnamefont {C.~D.}\ \bibnamefont {Hamley}}, \ and\ \bibinfo {author}
  {\bibfnamefont {M.~S.}\ \bibnamefont {Chapman}},\ }\href {\doibase
  10.1038/ncomms2179} {\bibfield  {journal} {\bibinfo  {journal} {Nat.
  Commun.}\ }\textbf {\bibinfo {volume} {3}} (\bibinfo {year} {2012}),\
  10.1038/ncomms2179}\BibitemShut {NoStop}%
\bibitem [{\citenamefont {Hoang}\ \emph {et~al.}(2013)\citenamefont {Hoang},
  \citenamefont {Gerving}, \citenamefont {Land}, \citenamefont {Anquez},
  \citenamefont {Hamley},\ and\ \citenamefont {Chapman}}]{Hoang13}%
  \BibitemOpen
  \bibfield  {author} {\bibinfo {author} {\bibfnamefont {T.~M.}\ \bibnamefont
  {Hoang}}, \bibinfo {author} {\bibfnamefont {C.~S.}\ \bibnamefont {Gerving}},
  \bibinfo {author} {\bibfnamefont {B.~J.}\ \bibnamefont {Land}}, \bibinfo
  {author} {\bibfnamefont {M.}~\bibnamefont {Anquez}}, \bibinfo {author}
  {\bibfnamefont {C.~D.}\ \bibnamefont {Hamley}}, \ and\ \bibinfo {author}
  {\bibfnamefont {M.~S.}\ \bibnamefont {Chapman}},\ }\href {\doibase
  10.1103/PhysRevLett.111.090403} {\bibfield  {journal} {\bibinfo  {journal}
  {Phys. Rev. Lett.}\ }\textbf {\bibinfo {volume} {111}},\ \bibinfo {pages}
  {090403} (\bibinfo {year} {2013})}\BibitemShut {NoStop}%
\bibitem [{\citenamefont {Goldstone}(1961)}]{Goldstone61}%
  \BibitemOpen
  \bibfield  {author} {\bibinfo {author} {\bibfnamefont {J.}~\bibnamefont
  {Goldstone}},\ }\href@noop {} {\bibfield  {journal} {\bibinfo  {journal}
  {Nuovo Cimento}\ }\textbf {\bibinfo {volume} {19}},\ \bibinfo {pages} {154}
  (\bibinfo {year} {1961})}\BibitemShut {NoStop}%
\bibitem [{\citenamefont {Damski}(2005)}]{Damski05}%
  \BibitemOpen
  \bibfield  {author} {\bibinfo {author} {\bibfnamefont {B.}~\bibnamefont
  {Damski}},\ }\href@noop {} {\bibfield  {journal} {\bibinfo  {journal} {Phys.
  Rev. Lett.}\ }\textbf {\bibinfo {volume} {95}},\ \bibinfo {pages} {035701}
  (\bibinfo {year} {2005})}\BibitemShut {NoStop}%
\bibitem [{\citenamefont {Farhi}\ \emph {et~al.}(2001)\citenamefont {Farhi},
  \citenamefont {Goldstone}, \citenamefont {Gutmann}, \citenamefont {Lapan},
  \citenamefont {Lundgren},\ and\ \citenamefont {Preda}}]{Farhi01}%
  \BibitemOpen
  \bibfield  {author} {\bibinfo {author} {\bibfnamefont {E.}~\bibnamefont
  {Farhi}}, \bibinfo {author} {\bibfnamefont {J.}~\bibnamefont {Goldstone}},
  \bibinfo {author} {\bibfnamefont {S.}~\bibnamefont {Gutmann}}, \bibinfo
  {author} {\bibfnamefont {J.}~\bibnamefont {Lapan}}, \bibinfo {author}
  {\bibfnamefont {A.}~\bibnamefont {Lundgren}}, \ and\ \bibinfo {author}
  {\bibfnamefont {D.}~\bibnamefont {Preda}},\ }\href@noop {} {\bibfield
  {journal} {\bibinfo  {journal} {Science}\ }\textbf {\bibinfo {volume}
  {292}},\ \bibinfo {pages} {472} (\bibinfo {year} {2001})}\BibitemShut
  {NoStop}%
\bibitem [{\citenamefont {Carr}(2010)}]{Carr10}%
  \BibitemOpen
  \bibfield  {author} {\bibinfo {author} {\bibfnamefont {L.}~\bibnamefont
  {Carr}},\ }\href@noop {} {\emph {\bibinfo {title} {Understanding Quantum
  Phase Transitions}}}\ (\bibinfo  {publisher} {CRC press},\ \bibinfo {year}
  {2010})\BibitemShut {NoStop}%
\bibitem [{\citenamefont {Pethick}\ and\ \citenamefont
  {Smith}(2002)}]{Pethick02}%
  \BibitemOpen
  \bibfield  {author} {\bibinfo {author} {\bibfnamefont {C.}~\bibnamefont
  {Pethick}}\ and\ \bibinfo {author} {\bibfnamefont {H.}~\bibnamefont
  {Smith}},\ }\href {http://books.google.com/books?id=iBk0G3\_5iIQC} {\emph
  {\bibinfo {title} {Bose-Einstein condensation in dilute gases}}}\ (\bibinfo
  {publisher} {Cambridge University Press},\ \bibinfo {year}
  {2002})\BibitemShut {NoStop}%
\end{thebibliography}%

\bibliographystyle{apsrev4-1}

\clearpage

\onecolumngrid

\begin{center}
\textbf{SUPPLEMENTAL INFORMATION}	
\end{center}

\vspace{10mm}

\twocolumngrid

In this supplemental information, we present the experimental methods used to prepare the system and determine the location of the critical point, and give additional details about the magnetic field ramps used to find the KZM scaling.
The relation between the spinor dynamical rate $c$ and the total number of atoms in the condensate is derived, and the methods behind the numerical simulations are explained. Finally, the plots showing the extraction of the scaling exponents are shown for every data set used in this study.

\section{Methods}

\textbf{System preparation}

The experiment starts by cooling $\sim$ $10^8$ $^{87}$Rb atoms in a magneto-optical trap (MOT).  
After 15~s, the number of atoms in the MOT reaches saturation, and a temporal dark MOT sequence is initiated. 
This procedure increases the spatial density of the trapped atoms as the MOT collapses along the repump axis, which increases the overlap with the $\sim$ 120~$\mu$m-waist optical trap, created by the focus of 10.6~$\mu$m light from a CO$_2$ laser. After 150~$\mu$s, there are 10-15 million atoms in the optical trap.
Evaporative cooling takes place for 3~s in the optical trap, during which a strong magnetic field gradient of $\sim$ 20~G/cm is applied along the laser axis, removing the $| F=1,m_F=\pm1\rangle$ atoms from the cloud, thus leaving only atoms in the $| F=1,m_F=0\rangle$ hyperfine ground state. 
For optimal cooling, the focus of the CO$_2$ laser must also be tightened, resulting in a beam waist of 20$~\mu$m. The evaporation leaves a nearly pure BEC with a temperature less than 100~nK, with no observable thermal component. 
Near the end of evaporation, a 850~nm beam with a waist of $\sim 20~\mu$m is gradually turned on, intersecting the CO$_2$ beam at their respective focuses. This confines the atoms in a nearly spherical trapping potential.

The atoms are initially held in a homogeneous magnetic field of 2~G, which corresponds to the polar ground state, so the system will not evolve.
For the experiments presented, the BEC contained $30(1) \times 10^3$ to $60(1) \times 10^3$ atoms.
These condensates have a Thomas-Fermi radius of $\sim$~5~$\mu$m and a spin healing length of $\sim$~10~$\mu$m, thus guaranteeing that the experiment is well described by the single mode approximation.
The experiment has also been performed in a single-focus, cigar-shaped trap formed by the focus of the CO$_2$ laser alone. 
In this trap, the condensate is no longer in the single mode approximation. Nevertheless, no spin domains were detected before the system crossed the threshold used to determine the freeze-out time, and there was no significant difference in the critical exponents.

After evolution, the populations in the three $m_F$ spin states are measured by releasing the trap and allowing the atoms to freely expand during 22~ms in a Stern-Gerlach magnetic field gradient. This gradient separates the three spin component clouds as they fall, preventing any spatial overlap. To count the three $m_F$ components, the atoms are probed for $200~\mu$s, and the fluorescence of the $m_F$ components is collected by a CCD camera.

\textbf{Critical point determination}

Determining the magnetic field $B_c$ corresponding to the critical point with high reliability and precision is necessary for a meaningful investigation of KZM scaling.
The procedure starts by initializing the system in the polar phase ground state at a high magnetic field, well above the critical point $(B_i\approx 2$~G~$ \gg B_{c})$, where $\rho_{0}$~=~1 is the ground state. 
The magnetic field is then quickly (2~ms) lowered to a fixed value $B_f$ near the expected critical field. 
If this final magnetic field value is above the critical point, the creation of $m_F=-1$ and $m_F=+1$ will be suppressed.
However, if the system finds itself in the broken-axisymmetry phase, spin-mixing dynamics will occur. 
These spin population dynamics are determined by how far the final magnetic field is from the critical point.

\begin{figure}[t]
	\begin{center}
		 \includegraphics[width=3.25in]{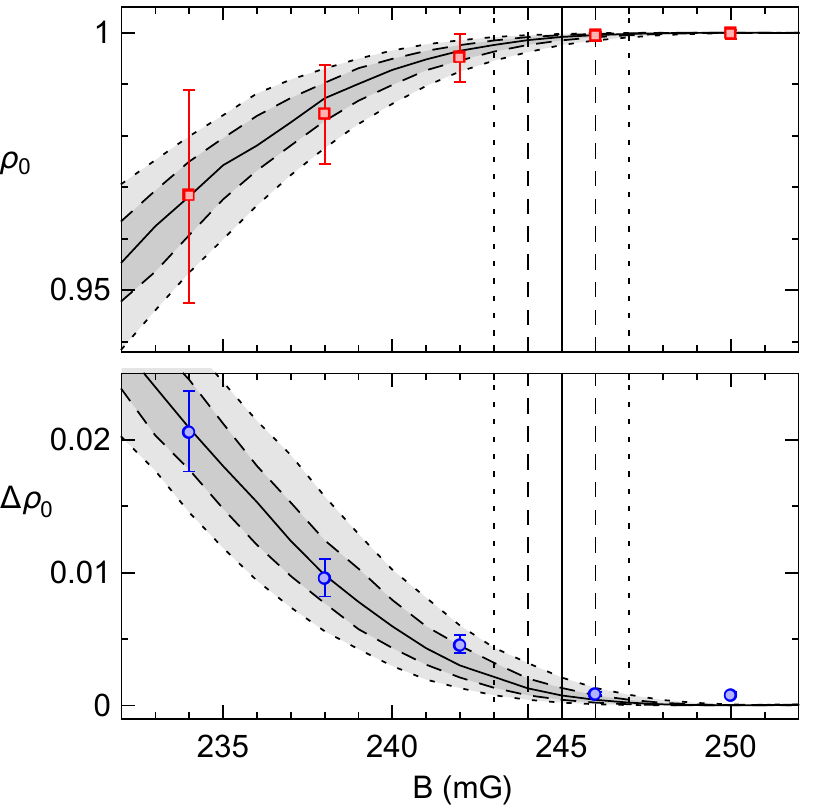}	
	\end{center}
    \caption[Evolution compared with range of simulations]{\textbf{Evolution compared with range of simulations.} 
	Measurements of $\rho_0$ (red squares) and $\Delta \rho_0$ (blue circles) after 500 ms of evolution at constant magnetic field (shown on the horizontal axis) following a fast (2 ms) drop from 2 G ($q\gg 2|c|$). In this plot, the data is compared with five simulations, represented by solid, dashed, and dotted curved lines. The solid line shows the simulation performed using a critical field of 245~mG, whereas the dashed and dotted lines represent simulations with critical fields of $\pm1$ and $\pm2$~mG, respectively. The vertical lines indicate the different critical fields used in the simulations.
    }
\label{fig:BCritScan}
\end{figure}

The system is allowed to evolve at the constant magnetic field $B_f$ for a time $t_{\mathrm{evol}}$ (ranging from 150~ms in the cross trap to 500~ms in the single focus trap) prior to measurement.
This process is repeated for several final fields and the spin population transfer is measured. The mean of the fractional population $\rho_0$ and its standard deviation $\Delta \rho_0$ are then plotted against the final value of the field.
An example of data and the simulation that is the best match (represented by a the solid black trace) is shown in \figref{fig:BCritScan}, where not only $\rho_0$ but also its standard deviation $\Delta\rho_0$ show good agreement with simulations. 

The precision of the determination of the critical point can be estimated by comparing the data with simulations performed using other critical fields, as illustrated in \figref{fig:BCritScan}. 
Using this technique, the uncertainty in the determination of the critical field can be as low as 1 to 2~mG.

\textbf{Magnetic field ramps}

Once the location of the critical point is precisely known, the next step is to ramp down the magnetic field through the critical point at different speeds, which is how the KZM scaling is determined.
The system is prepared in the polar ground state at high magnetic field ($B \approx 2$~G~$ \gg B_{c}$), and all the atoms are in the $m_F=0$ state. The system is first brought closer to the critical point by rapidly (2 ms) lowering the field such that $q = q_0 = 2.2|c| = 1.1q_c$. This value is still over the critical point, so the system is still very close to the polar ground state.

The system is then driven through the critical point by decreasing the magnetic field such that $q$ varies linearly as $q(t) = q_0 - t/\tau_Q  = q_0 (1 - t/t_r)$, where $t_r$ is the time it takes to ramp the field from $q=q_0$ to $q=0$. 
The experiment is repeated for several values of $t_r$. The fastest ramp uses $t_r=100$~ms, whereas the slowest ones last up to 4~s in the cross trap and 9~s in the single focus trap. 
After a freeze-out time $\hat{t}$ following the quantum phase transition, the system starts evolving. 
$\rho_0$ is measured multiple times at regular intervals during the ramp, and its mean and standard deviation $\Delta \rho_0$ are determined.

\section{Second-Order Phase Transition}

As expected from a second-order QPT, our system is characterized by a vanishing energy scale, namely the energy gap between the ground state and the first accessible excited state, which approaches zero at the critical point.
This vanishing energy scale results in critical slowing down, as observed in the divergence of the reaction time and correlation length of the system.
The reaction time is also known as the relaxation time, and gives the time scale at which the system can adiabatically follow a changing ground state, or return to its ground state after an excitation. An intuitive way to consider this phenomenon is that the reflexes of the system deteriorate around the critical point.
Similarly, the correlation length describes the scale on which the system can `heal' in space and collectively return to its ground state after an excitation.

The parameter from the Hamiltonian that is changed for the system to cross the critical point is the quadratic Zeeman energy $q \sim B^2$. The critical point takes place when $q = q_c = 2|c|$, where $c$ is the spinor dynamical rate, which essentially characterizes the energy from the spin interactions \cite{Lamacraft07}. 
In our experiment, we lower the magnetic field such that the QPT happens from the polar phase $(q > q_c)$ to the broken-axisymmetry phase $(q<q_c)$.

The expression for the energy gap in the broken-axisymmetry phase of our system is given by
\begin{equation}
\Delta =  \sqrt{q_c^2-q^2}.
\end{equation}
Close to the critical point, 
\begin{align}
\label{eq:EgapApprox}
\Delta &\approx \sqrt{2q_c(q_c-q)}\\
 & =  \Delta_0 \bigl| q_c - q \bigl| ^{1/2} ,
\end{align}
with $\Delta_0 = \sqrt{2q_c}$.
The energy gap approaching the critical point of a second-order phase transition is generically given by
\begin{equation}
	\Delta \sim |g_c-g|^{z\nu}.
\end{equation}
The product of the critical exponents $z$ and $\nu$ is $1/2$ for our system, corresponding to the mean field values of $z=1$ and $\nu=1/2$ \cite{Carr10}.

One of the key points behind the Kibble-Zurek mechanism is that the system cannot adiabatically follow the ground state when the ground state is changed too quickly compared to the reaction time. 
When the change occurs too fast, the evolution of the system switches from an adiabatic regime to an impulse regime where the system `freezes' with no evolution.  This is followed by the system `unfreezing,' returning to an adiabatic regime. 
This freezing of the dynamics happens when the reaction time diverges, which is the case when the energy gap between the ground state and the first excited state vanishes, as seen in our system. 
The transition from adiabatic to impulse and back to adiabatic regimes happens when the time scales characterizing the system's reaction time and the rate of change of the energy gap are comparable. This can be described by
\begin{equation}
\tau(\hat{t}) = \frac{1}{\Delta(\hat{t})} = \left. \frac{\Delta}{\mathrm{d}\Delta / \mathrm{d}t} \right|_{t=\hat{t}} \label{eq:comparableTimeScales}
\end{equation}
where $\hat{t}$ is the freeze-out time.
When $\dot{q} \propto 1/\tau_Q$ at the critical point, solving \eqnref{eq:comparableTimeScales} using the expression from \eqnref{eq:EgapApprox} and $z\nu = 1/2$ gives
\begin{equation}
\hat{t} =
 \left( z\nu \right)^\frac{1}{1+\nu z} \left(\frac{\tau_Q}{2q_c}\right) ^\frac{\nu z}{1+\nu z} = 
\left(\frac{\tau_Q}{8q_c}\right) ^{1/3} 
\sim \tau_Q ^{\ 1/3}.
\end{equation}

Introducing ${\tilde{q}(t)=q(t)/|c(t)|}$, where $c$ is the spinor dynamical rate of the system, we can define $\hat{q}$ as the change in $\tilde{q}$ from the crossing of the critical point to the return to the adiabatic regime.
After a similar derivation to the one for $\hat{t}$, we get
\begin{equation}
\hat{q} 
= 4^{\frac{-z\nu}{z\nu+1}} \left(\frac{\tilde{\tau}_Q}{z\nu}\right) ^\frac{-1}{1+\nu z}  
=\frac{1}{4^{2/3}} \tilde{\tau}_Q ^{\ -2/3} 
\sim \tilde{\tau}_Q ^{\ -2/3},
\end{equation}
where $\tilde{\tau}_Q \equiv  1/ \left| \dot{\tilde{q}}(t_c) \right|$ is a characteristic ramp time. 
These are the scaling exponents that are compared with the results from our measurements.

\section{Dependence of $c$ on the number of atoms in the condensate}

The spinor dynamical rate $c$ scales with the total number of atoms in the BEC. 
A brief derivation of the scaling is presented, and this theoretical prediction is then compared with experimental data. 

Within the Thomas-Fermi approximation \cite{Pethick02}, the density of atoms in the trap is given by
\begin{equation}
n_{\rm TF}(\vec{r})=\max \left[ \left( \frac{  \mu-U(\vec{r})}{c_0} \right) ,0 \right]~,
\label{eqn:TFdensity}
\end{equation}
where $\mu$ is the chemical potential, $U(\vec{r})$ is the trap potential, and $c_0 = 4\pi \hbar^2 a / m$ is the mean field density interaction strength. $a = (2a_2 + a_0)/3$ is an average scattering length which depends on the s-wave scattering lengths $ a_F $ for the collisions with total spin $F$, and $m$ is the mass of the atom.

In a harmonic potential with frequencies $\omega_i$,
\begin{equation}
n_{\rm TF}(\vec{r})=\frac{15N}{8\pi\Pi _i R_{i}}\max \left[ \left(1-\sum_{i=1}^{3}\frac{r_i^2}{R_{i}^2}\right),0\right],
\end{equation}\label{eqn:HarmonicTFdensity} 
where $N$ is the total number of atoms and the Thomas-Fermi radii $R_{i}$ are given by
\begin{equation}
R_{i}=\sqrt{\frac{2\mu }{m\omega_{i}^{2}}} \  .
\label{eqn:TFRadii}
\end{equation}
Using the normalization condition $\int n(\vec{r})\mathrm{d}^3r = N$, the chemical potential of the condensate can be calculated:
\begin{equation}
\mu = \left(\frac{15\hbar^{2}m^{1/2}}{2^{5/2}}N\bar{\omega}%
^{3}a\right) ^{2/5}, 
\label{eqn:ChemicalPotential}
\end{equation}
where $\bar{\omega}$ is the mean trap frequency.
From \eqnref{eqn:TFdensity}, we get a peak density $n_0 = \mu / c_0$. The value of $c$ is determined using $c \equiv c_2 N\int|\phi (\vec{r})|^4 d^3r$, where $c_2 = \frac{4\pi \hbar^2}{3m} (a_2 - a_0)$.
Solving the integral gives
\begin{equation}
\int | \phi (\vec{r})| ^{4}\mathrm{d}^{3}r = \frac{4}{7}\frac{\mu}{N c_0} .
\end{equation}
The factors of $N$ in front of and inside the integral in the expression for $c$ cancel out, which yields
\begin{equation}
c = \frac{4}{7} \frac{c_2}{c_0} \mu.
\end{equation}
The relative coupling strengths $c_0$ and $c_2$ only depend on the scattering lengths and the mass of the atom. 
In other words, the only dependence on the number of atoms $N$ in $c$ comes form the chemical potential $\mu$ from \eqnref{eqn:ChemicalPotential}:
\begin{equation}
c = \frac{4}{7} \frac{c_2}{c_0} \left(\frac{15\hbar^{2}m^{1/2}}{2^{5/2}}N\bar{\omega}%
^{3}a\right) ^{2/5}.
\end{equation}
This expression yields the scaling we are interested in: $c \propto N^{2/5}$.

\begin{figure}[h]
	\begin{center}
	\begin{minipage}{3.25in}
		 \includegraphics[scale=1]{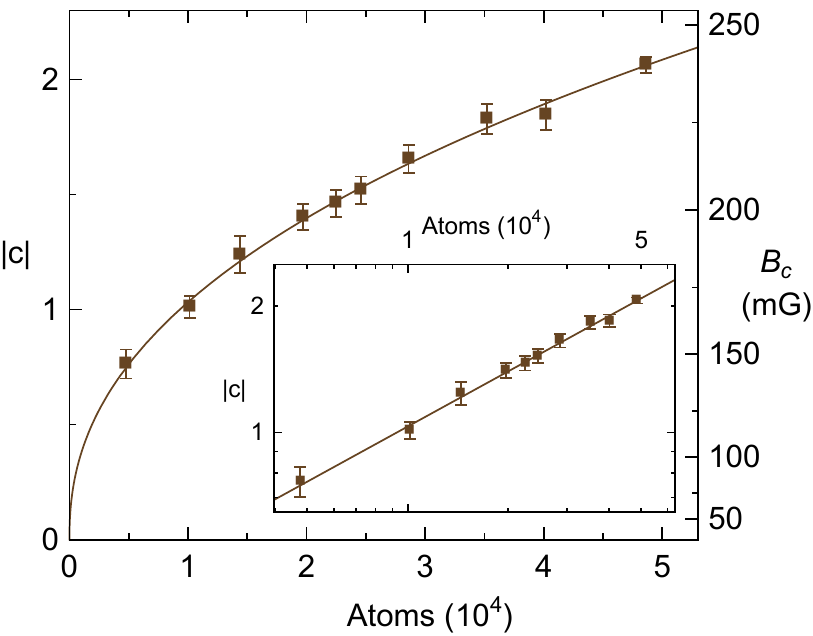}
		 \end{minipage}
	\end{center}
    \caption{\textbf{Spinor dynamical rate $|c|$ as a function of total number of atoms.} 
	The critical magnetic field was measured for different numbers of atoms in the condensate. The right axis shows the critical magnetic field. The inset shows the data in a log-log plot, exhibiting a clear power law dependence. A linear fit of the logarithm of the data yields an exponent of 0.44(2), in close agreement with the 2/5 predicted form theory.
    }
\label{fig:cVsN09-24-25}
\end{figure}

Using the technique described previously, the magnetic field corresponding to the critical point is determined from experimental data for different numbers of atoms in order to quantify the effect of atom loss on the spin dynamics.
The spinor dynamical rate $c$ is calculated from the magnetic field $B$. The quadratic Zeeman energy $q$ is given by $q = q_z B^2$, where $q_z \approx $ 71.6 Hz/G$^2$, and the critical point takes place at $q = q_c = 2|c|$.
\figref{fig:cVsN09-24-25} shows a measurement of $|c|$ versus the number of atoms in the condensate. The inset shows the data in a log-log plot, showing a clear power law dependence. 
The exponent determined from a linear fit of the logarithm of the data gives 0.44(2), which is within 10\% of the 0.4 exponent predicted by theory.

\section{Simulations}

The data is compared with mean field and quantum dynamical simulations. The mean field simulations are performed by numerically integrating the equations of motion of the order parameter $\psi ={{\left( {{\zeta }_{1}},{{\zeta }_{0}},{{\zeta }_{-1}} \right)}^{T}}$. 
However, our initial state is the polar ground state, where all the atoms are in the $m_F = 0$ energy level. 
This state does not evolve according to these equations \cite{Hamley12}, even in the broken-axisymmetry phase where the state is a hyperbolic fixed point. 
The evolution seen experimentally is due to quantum fluctuations.
In order to account for these quantum fluctuations in the mean field picture, a quasi-probability distribution is generated from the quantum noise of the initial Fock state $\left|0,N,0\right\rangle$ \cite{Hamley12}.

In the context of this study, this initial distribution essentially corresponds to a set of states slightly perturbed from the polar ground state.
The quadratic Zeeman energy $q$ is a parameter in the mean field equations of motion, so implementing a linear ramp in $q$, as in the experiment, is trivial. 
In order to account for atom loss, the spinor dynamical rate $c$ is updated with the number of atoms in the condensate such that $c \propto N^{2/5}$.
More details about the simulations and the generation of the quasi-probability distribution can be found in Ref. \cite{Hamley12}.

\textbf{Simulations in Ideal Conditions}

\begin{figure}[h]
	\centering
		 \includegraphics[scale=1]{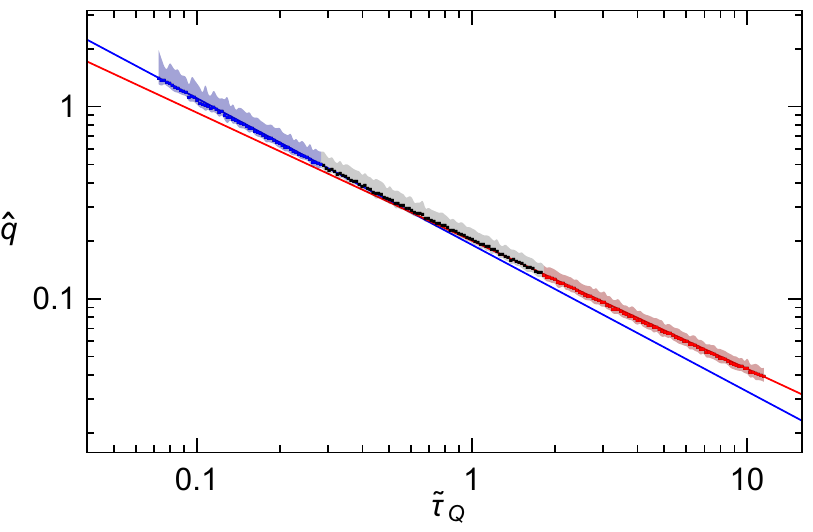}
   \caption{\textbf{Simulations in ideal conditions}. The ramps start from $8.7|c|$ (1~G), which is much higher than the critical point, as compared to the experiment. A fit of the faster ramps (blue) with $ 0.05 < \tilde{\tau}_Q < 0.28$ gives a scaling exponents of $-0.76(5)$. The slowest ramps (red) with $ \tilde{\tau}_Q > 1.83$ yield scaling exponents of $-0.67(2)$.}
\label{fig:simNoLoss1M1G}
\end{figure}

We begin by presenting simulations performed in ideal conditions: infinite condensate lifetime, asymptotically long ramps starting at a magnetic field much higher than the critical magnetic field, and a large number of atoms.
We simulate ramps starting at a higher magnetic field ($q = 4.3 q_c$) than in the experiment ($q = 1.1 q_c$), with a range of ramp times reaching well beyond the longest ramps used in the experiment. The output of the simulation can be seen in \figref{fig:simNoLoss1M1G}.
Even though the plot shows a slight curvature, good quality power law fits can be preformed in different ranges of ramp speeds.
The range of slow ramps shown in red in \figref{fig:simNoLoss1M1G} satisfies the asymptotic settings used to derive the KZM scaling: slow ramps starting from a high magnetic field with a large number of atoms.
In these ideal conditions, the scaling exponent extracted from simulations matches the value of $-2/3$ predicted by the KZM, which was also found using numerics by Damski and Zurek in Ref. \cite{Damski07}.

The reason the faster ramps deviate slightly from the KZM is the following: the derivation of scaling exponents described earlier assumes that the system is close to the critical point, which is where the universal critical exponents in the expressions for the energy gap appear. 
In our case, we approximate the energy gap $\Delta$ in the broken-axisymmetry phase ($q<q_c$) by
\begin{equation}
\Delta =  \sqrt{q_c^2-q^2} \approx \sqrt{2q_c(q_c-q)},
\end{equation}
and in the polar phase ($q > q_c$) by
\begin{equation}
\Delta =  2\sqrt{q(q-q_c)} \approx 2\sqrt{q_c(q-q_c)}.
\end{equation}
These approximations are only valid when $q \approx q_c$.
This means that for the derivation of the scaling exponents to be valid, the system must be driven slow enough such that the dynamics cease and resume their adiabatic evolution where the energy gap approximation is appropriate.

\textbf{Similar Ramps as in the Experiment}

Limited trap lifetimes set a limit regarding how high we can start the magnetic field ramps. Ideally, one would want to start a ramp at a field where the spin interactions are completely dominated by the quadratic Zeeman energy term from the Hamiltonian, and slowly ramp the magnetic field down towards the critical point. The magnetic field where we prepare the system in the polar ground state essentially suppresses any spin dynamics. 
However, if we were to lower the field using linear ramps in $q$ starting from that value, we would be constrained to using ramps that reach the critical point without significant losses, which would result in a fast rate of change of the field at the critical point. 
In the lab, a compromise is reached by starting with a fast drop from 2~G to a value close to the critical field, but still sufficiently above it so as to prevent any spin-mixing dynamics.

\begin{figure}[h]
	\centering
		 \includegraphics[scale=1]{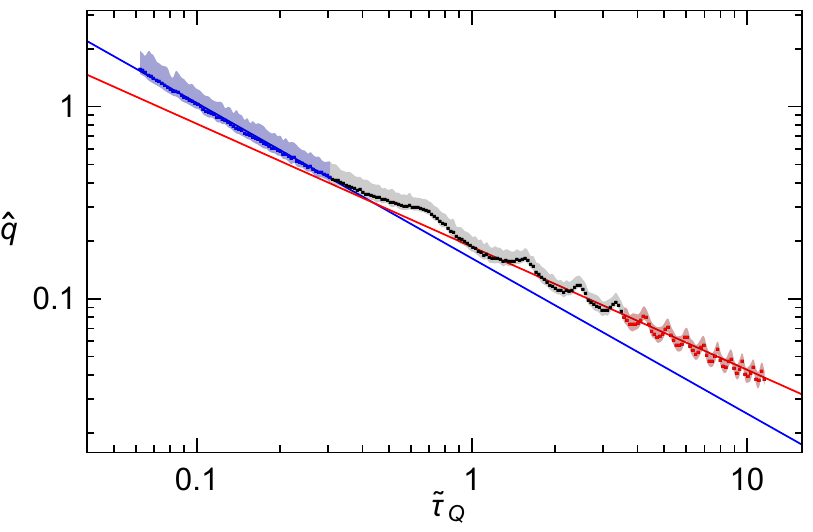}
    \caption{\textbf{Loss-less simulations with similar ramps as in the experiment}. The ramps start from $q=1.1q_c$ after a fast 2~ms drop from 2~G, as in the experiment.    
    The fits of the linear region (blue) with $ 0.06 < \tilde{\tau}_Q < 0.31 $ gives a scaling exponent of $-0.81(3)$. 
    A fit for the slower ramps (red) gives an exponent of $-0.64(3)$ for $ \tilde{\tau}_Q > 3.58 $.}
\label{fig:simNoLoss1M500mG}
\end{figure}

\figref{fig:simNoLoss1M500mG} shows a loss-less simulation using a similar initial fast drop as in the experiment. The faster ramps show a clear power law dependence in $\hat{q}$, followed by an oscillatory behavior as the ramp times increase. 
For the faster ramps, corresponding to the region plotted in blue, the scaling exponent is $-0.81(3)$.

\begin{figure}[h]
	\begin{center}
	\begin{minipage}{3.25in}
\begin{tabularx}{\linewidth}{ X  X } 
		 \includegraphics[scale=1]{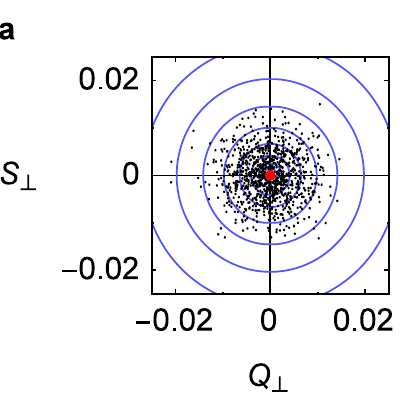}
\hfill 
		 \includegraphics[scale=1]{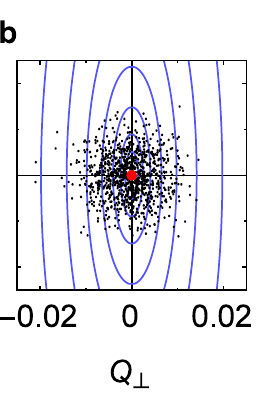}
\end{tabularx}		  
	\end{minipage}
	\end{center}
    \caption{\textbf{Initial distributions of states and energy contours before and after the initial quench}. These figures illustrate the effect of the 2~ms quench on the energy contours in the polar phase space. The figures shows energy contours at magnetic fields of 2~G (a) and 500~mG (b). The critical magnetic field is 480~mG.
}
\label{fig:InitialDistributionsBeforeAndAfterQuench}
\end{figure}

The oscillations in \figref{fig:simNoLoss1M500mG} are caused by the 2~ms fast drop that lowers the magnetic field from 2~G to a field corresponding to $q = 1.1q_c$. 
Even though $q = 1.1q_c$ is higher than the critical field and the system is still in the polar phase, the energy contours in the spin-nematic phase space around the polar ground state suddenly change from circles to ellipses, as illustrated in \figref{fig:InitialDistributionsBeforeAndAfterQuench}.

Before the quench, the initial distribution that corresponds to a slightly perturbed polar ground state is circularly symmetric and precesses around the ground state on the high field circular energy contours, as seen in \figref{fig:InitialDistributionsBeforeAndAfterQuench}a. 
Following the initial fast quench, the energy contours become ellipses, shown in \figref{fig:InitialDistributionsBeforeAndAfterQuench}b. As the initially circular distribution precesses around the ground state, its shape morphs back and forth from circular to elliptical until crossing the critical point and reaching the transition to the broken-axisymmetry phase. 
The ground state, which lay on the pole of the spin-nematic sphere in the polar phase, drifts down the (degenerate) sides of the sphere along the $S_{\bot}$ axis, which happens to be the major axis of the elliptical energy contours around the polar ground state.
The reaction time of the system depends on the shape and orientation of the distribution as the system enters the broken-axisymmetry phase.

\begin{figure}[h]
	\begin{center}
	\begin{minipage}{3.25in}
\begin{tabularx}{\linewidth}{ X  X } 
		 \includegraphics[scale=1]{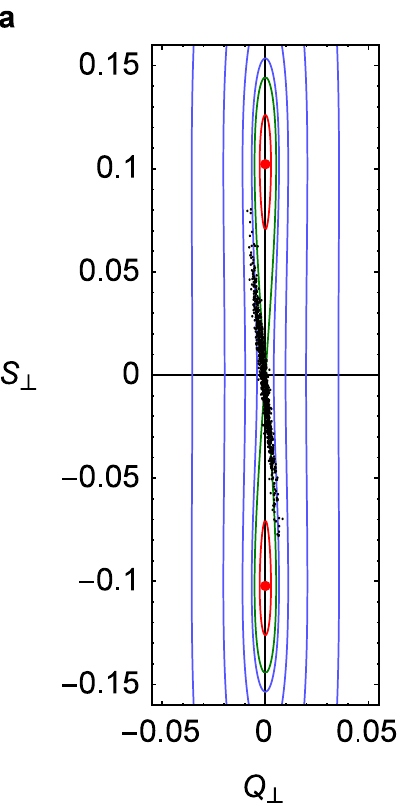}
\hfill 
		 \includegraphics[scale=1]{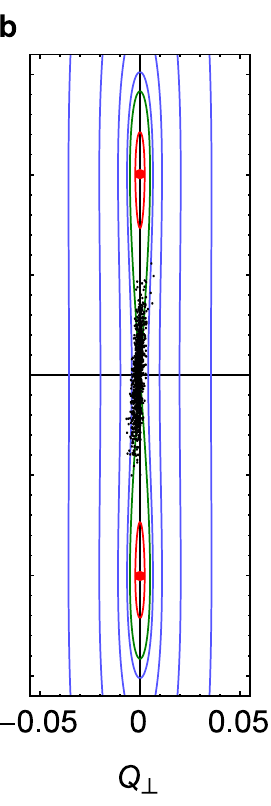}
\end{tabularx}		  
	\end{minipage}
	\end{center}
    \caption{\textbf{Polar phase space right after the critical point ($\tilde{q}$~=~1.99)}. The polar phase space shows the separatrix in green, the phase winding contours in  blue, and a red closed orbit, centered around the ground state represented by a red dot. The black dots represent 1000 samples initialized with a Gaussian distribution (for ${4 \times 10^4}$ atoms) around the pole. (a) shows the distribution during a 1.66~s ramp from 500~mG to 0~mG, and (b) during a 2~s ramp with a critical field $B_c =$~480~mG. Due to the precession of the distribution on the elliptical phase winding energy contours above the critical point, the distribution will be aligned along the diverging arm of the separatrix in (a), and the converging arm in (b) as the system crosses the critical point. This results in oscillations in the $\hat{q}$ plots for ramps beginning at $\tilde{q}$~=~2.2, as in \figref{fig:simNoLoss1M500mG}}. 
\label{fig:distCombined}
\end{figure}

At that time, the top pole of the spin-nematic sphere is a hyperbolic fixed point. 
As illustrated in \figref{fig:distCombined}, a separatrix (green) marks the boundary between the closed orbits (red) and the phase-winding orbits (blue). 
A state situated in the vicinity of the separatrix will evolve parallel to it, clockwise around the ground state. This means that states in the neighborhood of the pole will tend to evolve towards or away from it, depending in which quadrant of the polar phase they are located. The quadrants where $S_\bot$ and $Q_\bot$ have the same sign contain the converging branch of the separatrix, while the other two quadrants contain the diverging branch.

If the distribution is in a stretched elliptical shape aligned along the diverging branch of the separatrix, as in \figref{fig:distCombined}a, most states in the distribution will be able to follow energy contours leading away from the pole, shortening the mean reaction time of the system. 
However, if the distribution is in a circular shape, or as an ellipse aligned with the converging branch of the separatrix at the critical time, as illustrated in \figref{fig:distCombined}b, most states will be evolving around energy contours bringing them back to the pole and away from the broken-axisymmetry ground state, thus delaying the evolution and increasing the freeze-out time.

As shown earlier, starting the ramps at a higher magnetic field eliminates the oscillations in the $\hat{q}$ plots, which are only observed when the ramps start at the same field as the experiment.
The exponents extracted for different initial magnetic fields are included in \tableref{table:qHatSemiClassicalNoLossLongRamps}.

\begin{table}[h!]
	\caption[Summary of exponents from loss-less simulations]{\textbf{Summary of exponents from loss-less simulations.} This table summarizes the scaling exponents for $\hat{q}$ extracted from fits of loss-less simulations.
	}
\vspace{0.5cm}
	\centering
		\begin{tabular}{l l l l l}
		\hline
		\multicolumn{1}{ l }{Initial ramp field} &
		\multicolumn{1}{ l }{\qquad \qquad}  &
		\multicolumn{1}{ l }{Fast ramps}  &
		\multicolumn{1}{ l }{\qquad \qquad}  &
		\multicolumn{1}{ l }{Slow ramps} \\
		\hline
		1~G &  & $-0.76(4)$ &   & $-0.67(2)$ \\
		0.5~G &   & $-0.81(3)$ &   & $-0.64(3)$  \\
		\hline
		\end{tabular}

\label{table:qHatSemiClassicalNoLossLongRamps}
\end{table}

\textbf{Simulations Including Atom Loss}

Despite clear power law fits shown by the data for the short ramps, the values of 
$\hat{q}$ depart from the power law for the slowest ramps. 
When atom loss is included for simulations, the number of atoms is modeled using a double exponential $N(t) = \frac{N(0)}{2} (e^{-t / \tau_1} + e^{-t / \tau_2})$, with lifetimes determined from the typical behavior of our experiment. 
When simulating the conditions in the cross trap, the lifetimes are $ \tau_1 = $ 1~s and $ \tau_2 = $ 4.5~s. 
In the single focus trap, which has a much longer lifetime, we use $ \tau_1 = $ 15~s and $ \tau_2 = $ 30~s.
The effect of atom loss is included in the analysis by changing the spinor dynamical rate $c$, in the same fashion as for the data analysis.
Unless specified otherwise, the threshold used to determine the return to the adiabatic regime is $\rho_0 = 0.99$, and the ramp times $t_r$ range from 100~ms to 5~s.

\begin{figure}[h]
	\centering
		 \includegraphics[scale=1]{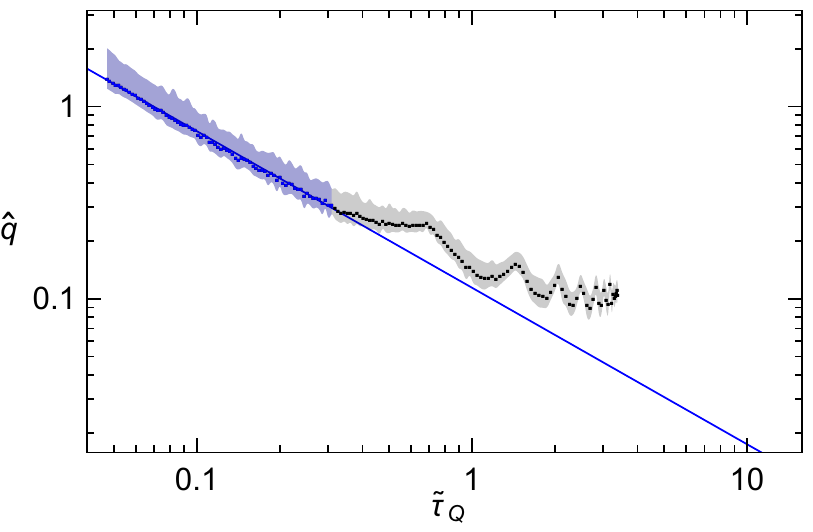}
    \caption{\textbf{Simulations including atoms loss}. Similarly as in the experiment, the ramps begin at $q = 1.1q_c$ and the initial number of atoms is $4 \times 10^4$ atoms. Fits of the linear region (blue) with $ 0.05 < \tilde{\tau}_Q < 0.31 $ give a scaling exponent of $-0.80(3)$.}
\label{fig:simsWithLoss}
\end{figure}

In order to emulate the behavior of our experiment, the simulations use the same magnetic field ramps as for the data. 
The linear $q$ ramps start from $q=1.1q_c$, following a fast drop from 2~G. 
Unlike the loss-less simulations presented earlier, we are constrained by limited trap lifetimes, which means we cannot initialize the ramps at higher magnetic fields than the experiment or test very slow ramps. A simulation including atom loss is shown in \figref{fig:simsWithLoss}.

\textbf{Comparison of Semi-Classical and Quantum Simulations}

The simulations presented so far were semi-classical, using the mean field dynamical equations along with an initial distribution of states that mimics the quantum fluctuations due to the finite number of atoms. 
Semi-classical simulations are used instead of quantum simulations simply because the former are orders of magnitude faster to run.

\begin{figure}[h!]
	\centering
		 \includegraphics[scale=1]{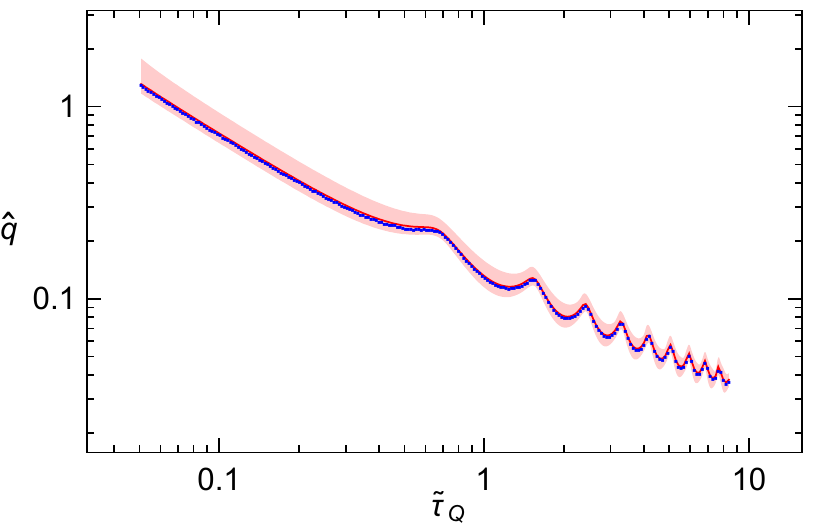}
	\caption{\textbf{Comparison of semi-classical and quantum simulations}. The outputs from loss-less semi-classical and quantum simulations for $4 \times 10^4$ atoms are compared. The ramps start at 500~mG, with a critical magnetic field set at $B_c = 480$~mG, which explains the oscillations for longer ramps. The solid red line is the output of the quantum simulation, and the shaded region represents one standard deviation. The blue dots show the output of the semi-classical simulation, performed with 1000 samples.}
\label{fig:qHatSemiClassicalVsQuantum}
\end{figure}

One may question the validity of using semi-classical simulations over quantum simulations for the following reason. 
The KZM theory insists on the fact that the energy gap between the ground state and the first excited state vanishes at the critical point, thus preventing the crossing of the critical point from occurring adiabatically. However, this claim is only strictly valid in the thermodynamic limit, for an asymptotically large numbers of atoms. With $\sim 4 \times 10^4$ atoms, the energy gap is actually non-zero because of the finite size of the condensate. 
Testing the effect of this non-vanishing energy gap requires the quantum version of our simulations.
A comparison of semi-classical and quantum simulations using $4 \times 10^4$ atoms is shown in \figref{fig:qHatSemiClassicalVsQuantum}. The overlap of the two outputs is excellent, thus justifying the interchangeable use of semi-classical or quantum simulations.

\onecolumngrid
\newpage
\begin{center}
\section{Measurements of $\rho_0$ during magnetic field ramps through the critical point} 
\end{center}
\twocolumngrid

The data shown in the plots below was taken during ramps lowering the field from 500~mG to 0~mG such that $q$ decreased linearly, with a critical magnetic field $B_c =$~480~mG. Each point and its associated error bars correspond to the mean of and standard deviation of 10 measurements at a given time. The solid red line that connects the data points illustrates how the time $t_{\mathrm{th}}$ when $\rho_0$ crosses the threshold is determined. Graphically, $t_{\mathrm{th}}$ corresponds to the $x$-coordinate of the intersection between the line connecting the points and the horizontal red line, which shows the threshold of $\rho_0$~=~0.99. The time $t_{\mathrm{th}}$ is represented by the vertical dashed line. Similarly, the upper and lower ends of the error bars are connected by gray dashed lines. The edges of the darker  shaded region around the vertical dashed line, which show the uncertainty in $t_{\mathrm{th}}$, are determined by the intersection between the horizontal red line and the gray dashed lines. The lighter shaded regions show the uncertainty added by the finite time step between the measured evolution times. The vertical red line shows the time $t_{c}$ of the critical point, and the shaded area is the uncertainty in its position. 
The values and errors of $\hat{t}$ and $\hat{q}$ are shown in the insets.

\onecolumngrid

\begin{figure}[h]
	\centering
	\begin{minipage}{3.25in}
		 \includegraphics[scale=1]{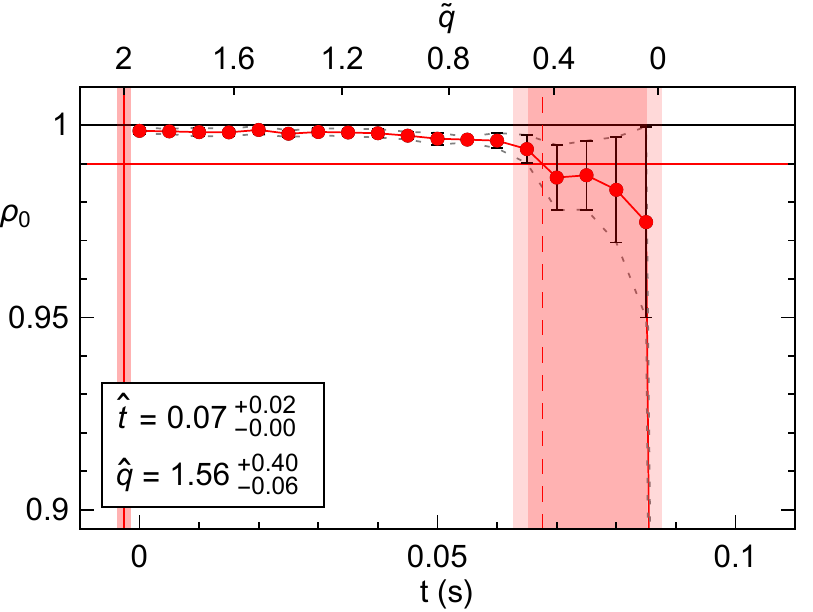}
    \caption{$t_r$ = 0.1 s, $\tilde{\tau}_Q$ = 0.05 s}
\label{fig:tr100ms}
	\end{minipage}
\hfill
	\begin{minipage}{3.25in}
		 \includegraphics[scale=1]{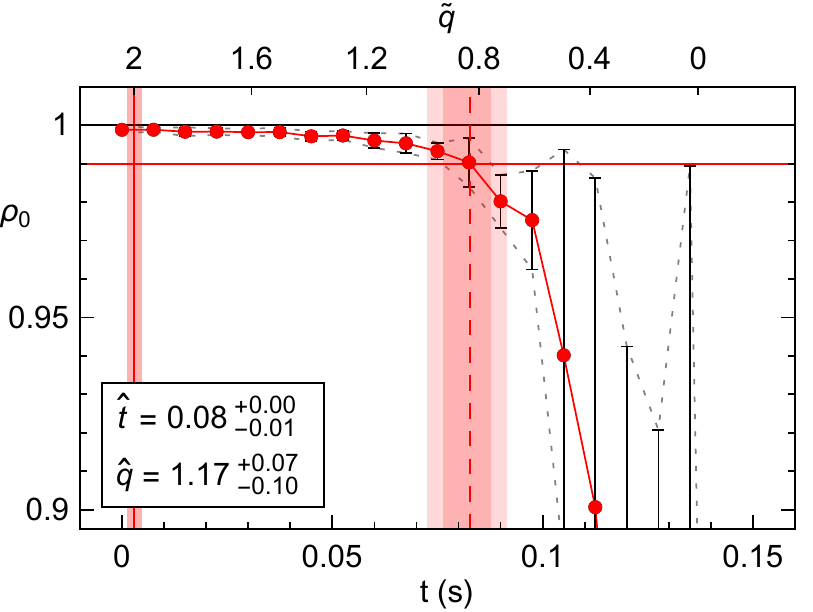}
    \caption{$t_r$ = 0.15 s, $\tilde{\tau}_Q$ = 0.07 s}
\label{fig:tr150ms}
	\end{minipage}
\end{figure}

\begin{figure}[h]
	\centering
	\begin{minipage}{3.25in}
		 \includegraphics[scale=1]{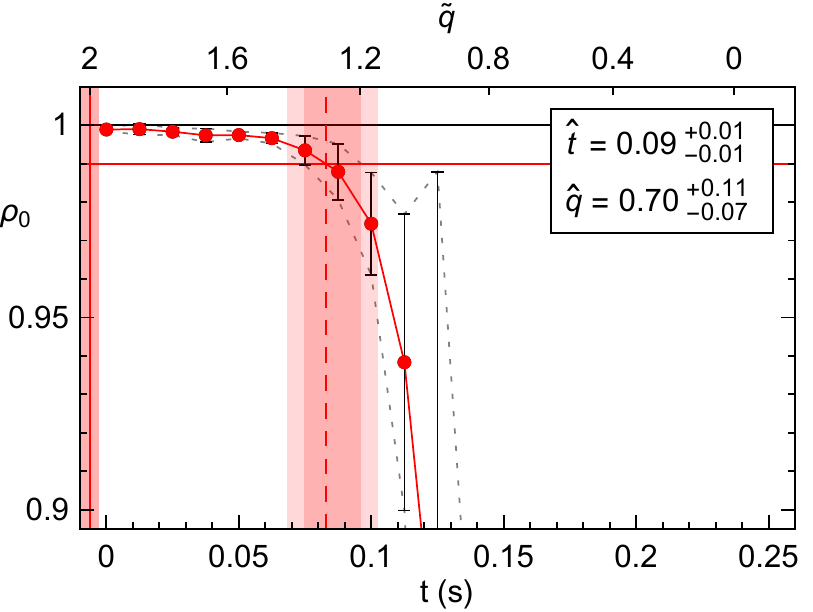}
    \caption{$t_r$ = 0.25 s, $\tilde{\tau}_Q$ = 0.13 s}
\label{fig:tr250ms}
	\end{minipage}
\hfill
	\begin{minipage}{3.25in}
		 \includegraphics[scale=1]{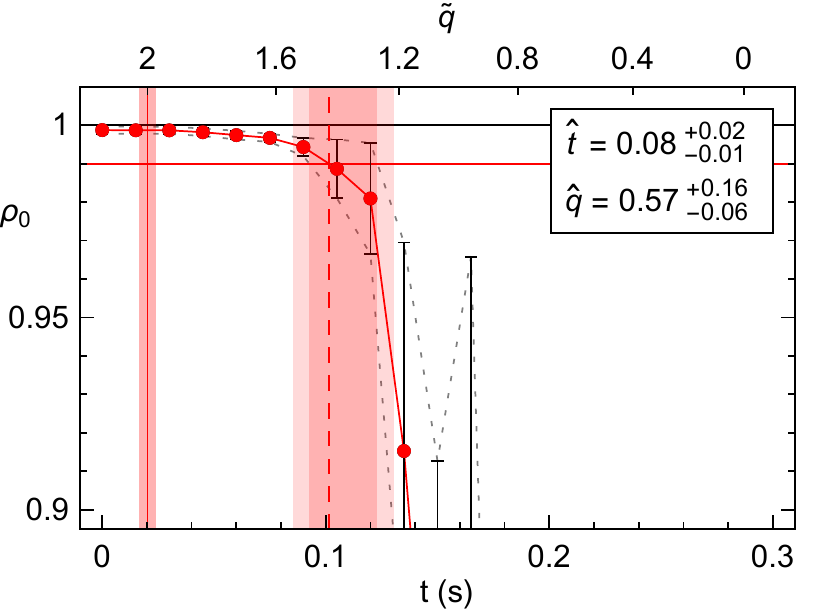}
    \caption{$t_r$ = 0.3 s, $\tilde{\tau}_Q$ = 0.15 s}
\label{fig:tr300ms}
	\end{minipage}
\end{figure}

\begin{figure}[h]
	\centering
	\begin{minipage}{3.25in}
		 \includegraphics[scale=1]{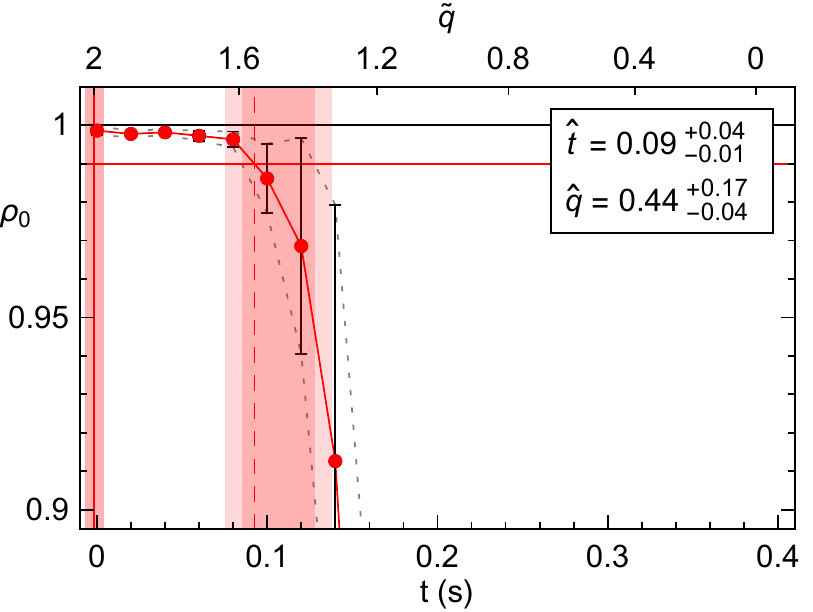}
    \caption{$t_r$ = 0.4 s, $\tilde{\tau}_Q$ = 0.22 s}
\label{fig:tr400ms}
	\end{minipage}
\hfill
	\begin{minipage}{3.25in}
		 \includegraphics[scale=1]{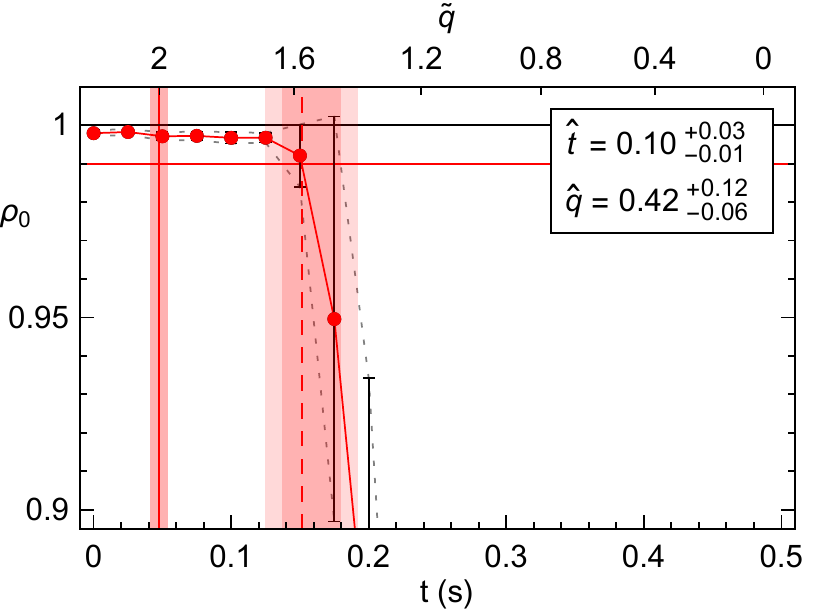}
    \caption{$t_r$ = 0.5 s, $\tilde{\tau}_Q$ = 0.25 s}
\label{fig:tr500ms}
	\end{minipage}
\end{figure}

\begin{figure}[h]
	\centering
	\begin{minipage}{3.25in}
		 \includegraphics[scale=1]{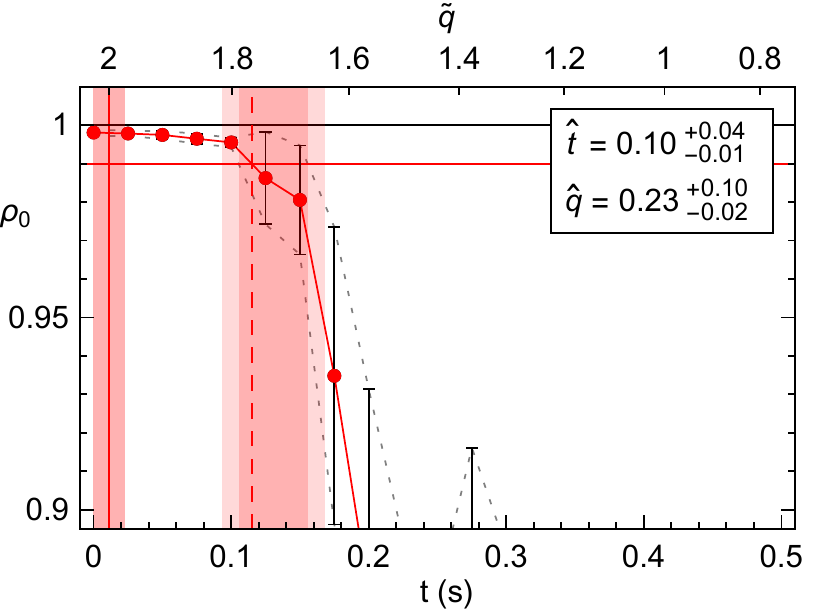}
    \caption{$t_r$ = 0.75 s, $\tilde{\tau}_Q$ = 0.46 s}
\label{fig:tr750ms}
	\end{minipage}
\hfill
	\begin{minipage}{3.25in}
		 \includegraphics[scale=1]{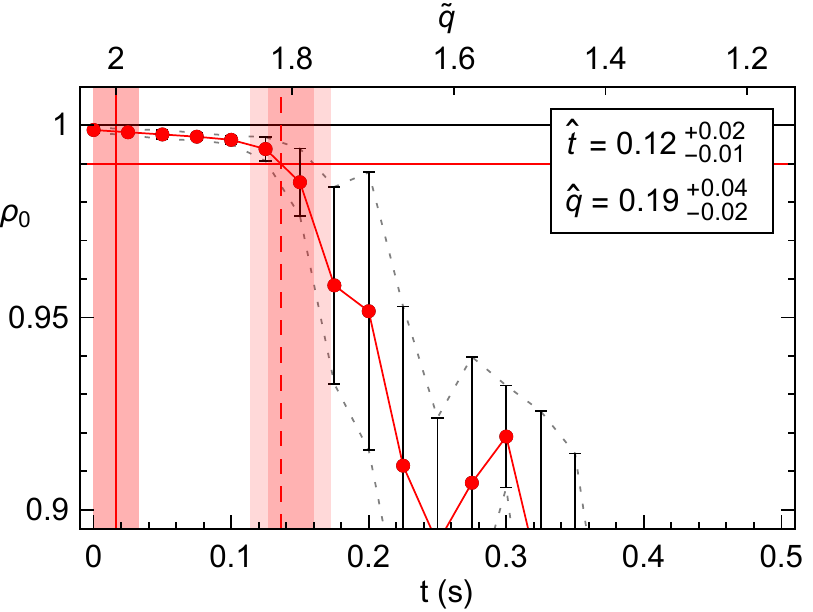}
    \caption{$t_r$ = 1 s, $\tilde{\tau}_Q$ = 0.67 s}
\label{fig:tr1000ms}
	\end{minipage}
\end{figure}

\begin{figure}[h]
	\centering
	\begin{minipage}{3.25in}
		 \includegraphics[scale=1]{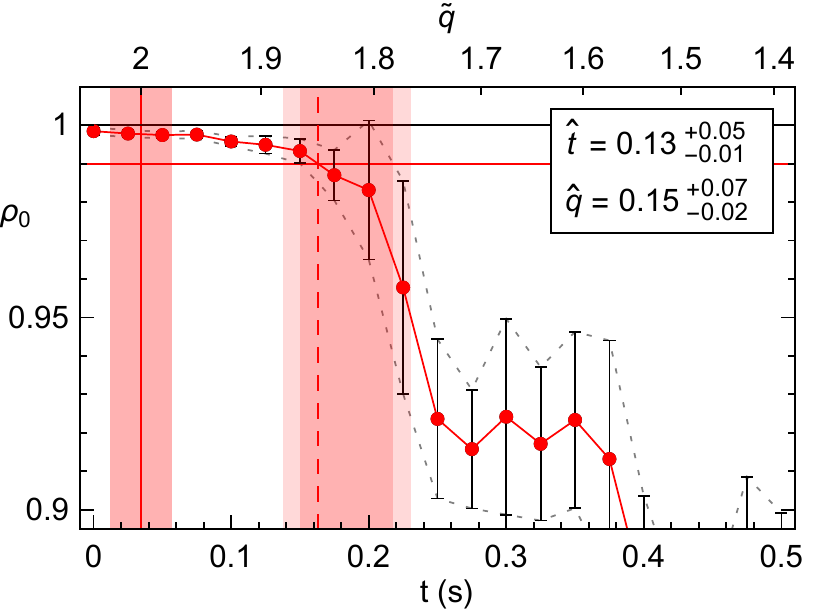}
    \caption{$t_r$ = 1.25 s, $\tilde{\tau}_Q$ = 0.90 s}
\label{fig:tr1250ms}
	\end{minipage}
\hfill
	\begin{minipage}{3.25in}
		 \includegraphics[scale=1]{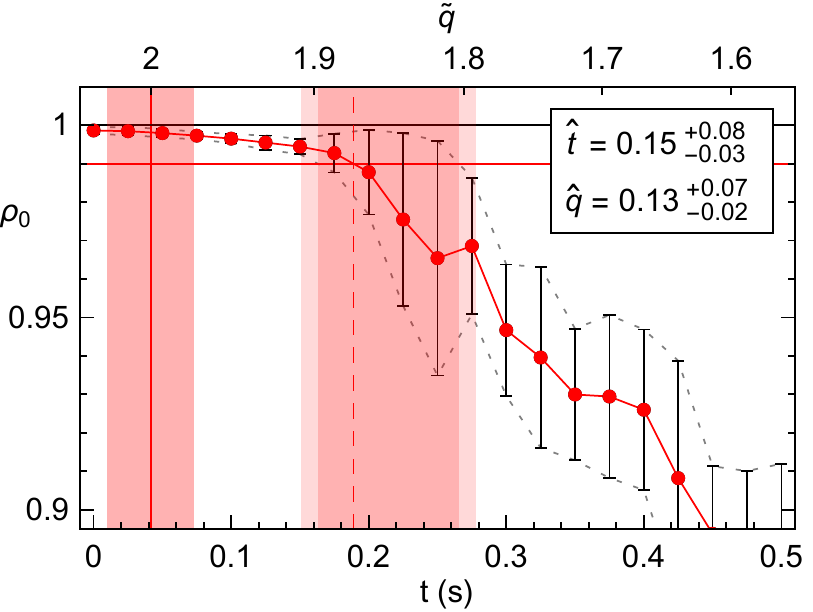}
    \caption{$t_r$ = 1.5 s, $\tilde{\tau}_Q$ = 1.24 s}
\label{fig:tr1500ms}
	\end{minipage}
\end{figure}

\begin{figure}[h]
	\centering
	\begin{minipage}{3.25in}
		 \includegraphics[scale=1]{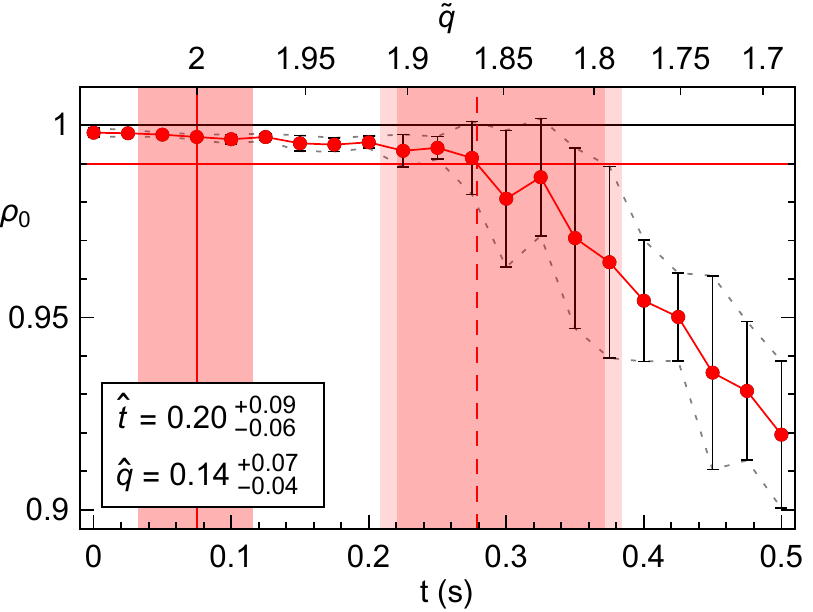}
    \caption{$t_r$ = 1.75 s, $\tilde{\tau}_Q$ = 1.64 s}
\label{fig:tr1750ms}
	\end{minipage}
\hfill
	\begin{minipage}{3.25in}
		 \includegraphics[scale=1]{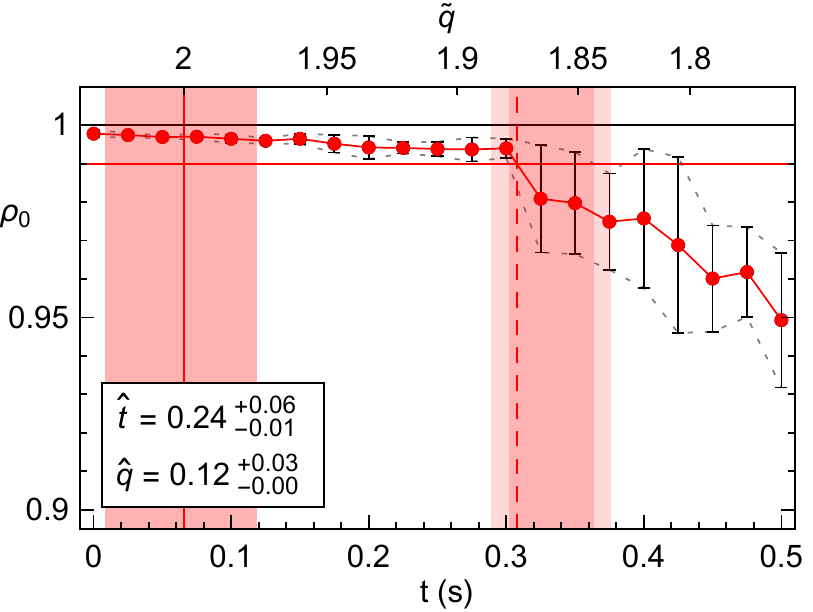}
    \caption{$t_r$ = 2 s, $\tilde{\tau}_Q$ = 2.18 s}
\label{fig:tr2000ms}
	\end{minipage}
\end{figure}

\begin{figure}[h]
	\centering
	\begin{minipage}{3.25in}
		 \includegraphics[scale=1]{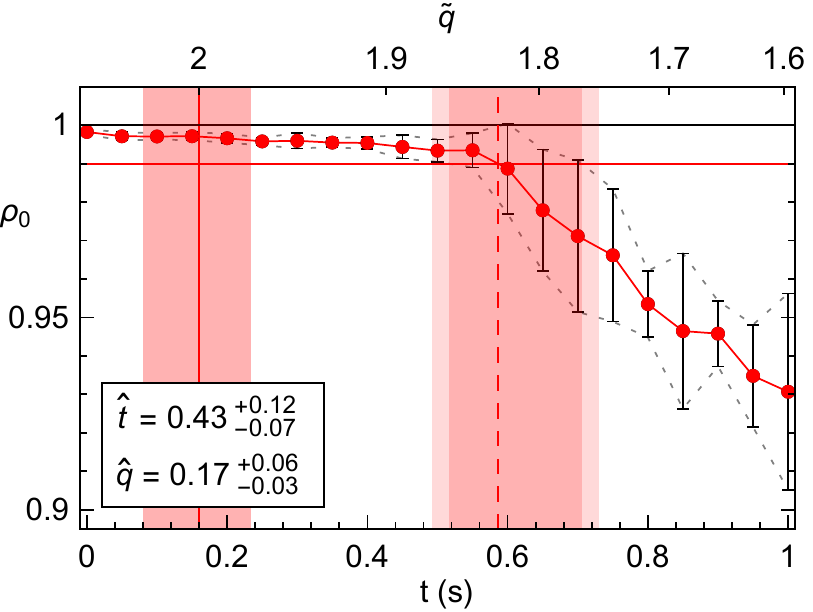}
    \caption{$t_r$ = 2.5 s, $\tilde{\tau}_Q$ = 3.02 s}
\label{fig:tr2500ms}
	\end{minipage}
\hfill
	\begin{minipage}{3.25in}
		 \includegraphics[scale=1]{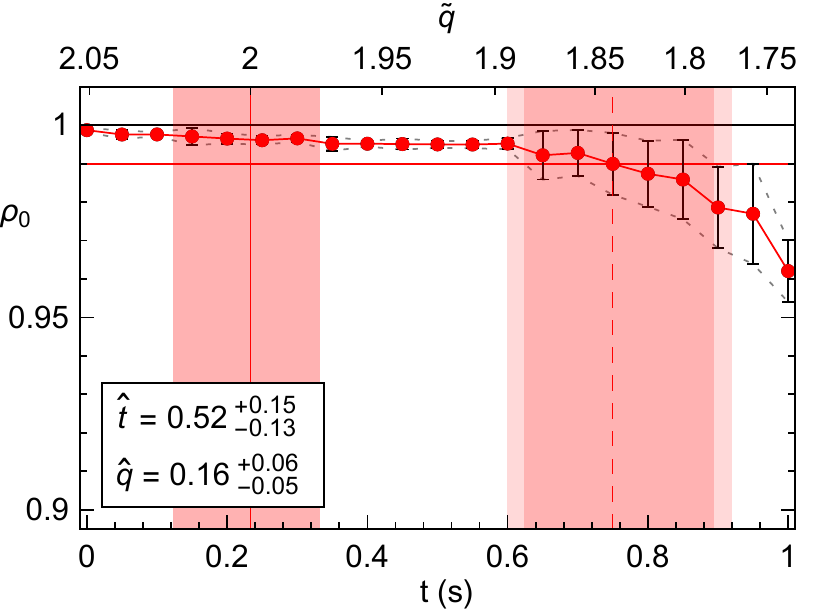}
    \caption{$t_r$ = 3 s, $\tilde{\tau}_Q$ = 4.10 s}
\label{fig:tr3000ms}
	\end{minipage}
\end{figure}

\clearpage

\onecolumngrid

\begin{center}

\vspace{1cm}
\section{Plots of $\hat{q}$ for all data sets}

\end{center}

\twocolumngrid

In the following plots, $\hat{q}$ is plotted with respect to the characteristic magnetic field ramp time $\tilde{\tau}_Q$, given by the inverse of the rate of change of $\tilde{q}$ at the critical point. 
The vertical error bars are found by combining the uncertainties in the determination of the time $t_c$ the system crosses the critical point and in the time $t_\mathrm{th}$ the system crosses the $\rho_0$ and $\Delta\rho_0$ thresholds. 
The characteristic ramp times $\tilde{\tau}_Q$ integrate atom loss by including the changing value of $c$, whose initial value is calculated from the measurement of the magnetic field at the critical point. The error in the determination of the critical magnetic field therefore induces an uncertainty in $\tilde{\tau}_Q$ represented by the horizontal error bars.
For every data point, a simulation is performed with the same experimental parameters (ramp time, number of atoms, and initial value of $c$) and their corresponding errors . For clarity, the simulations are plotted by interpolating between the output points as a gray dashed line, with a grey envelope displaying the error calculated using the same method as for the data.
The insets show the data and simulations plotted in a log-log plot. Linear fits to the log of the data and simulations give the critical exponents. 
For the fits, the points represented by empty markers are not used, which restricts the fitting to the linear regions of the data, indicated by solid square markers. 
For each of the four data sets shown, the red plots show the values of $\hat{q}$ determined by measuring $\rho_0$ and using a threshold of $\rho_0 = $ 0.99 to determine $t_\mathrm{th}$. The blue plots use the standard deviation $\Delta\rho_0$ and a threshold of $\Delta\rho_0 = $ 0.01.

\onecolumngrid
\vspace{0.5cm}

\begin{figure}[h]
	\centering
	\begin{minipage}{3.25in}
		 \includegraphics[width=3.25in]{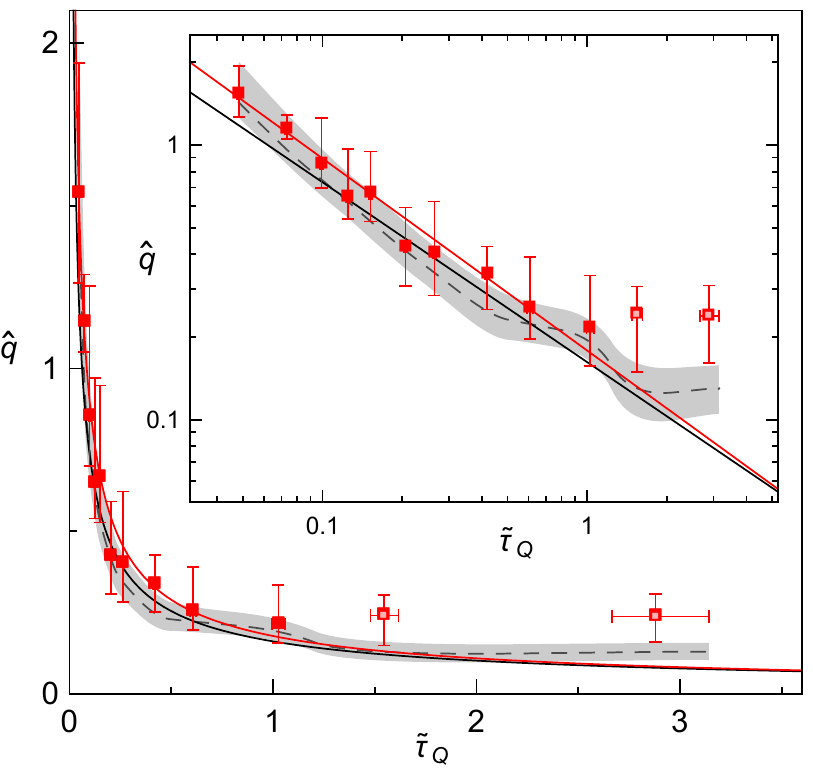}
    \caption{\textbf{Data set \#1, $\mathbf{\rho_0}$ threshold}. Plot range: 0.048 $ < \tilde{\tau}_Q < $ 2.88. Fit range: 0.048 $ < \tilde{\tau}_Q < $ 1.03. Critical exponents from fits: data: -0.70(9), simulations: -0.65(7).
   }
\label{fig:0127qhatrho0}
	\end{minipage}
\hfill
	\begin{minipage}{3.25in}
		 \includegraphics[scale=1]{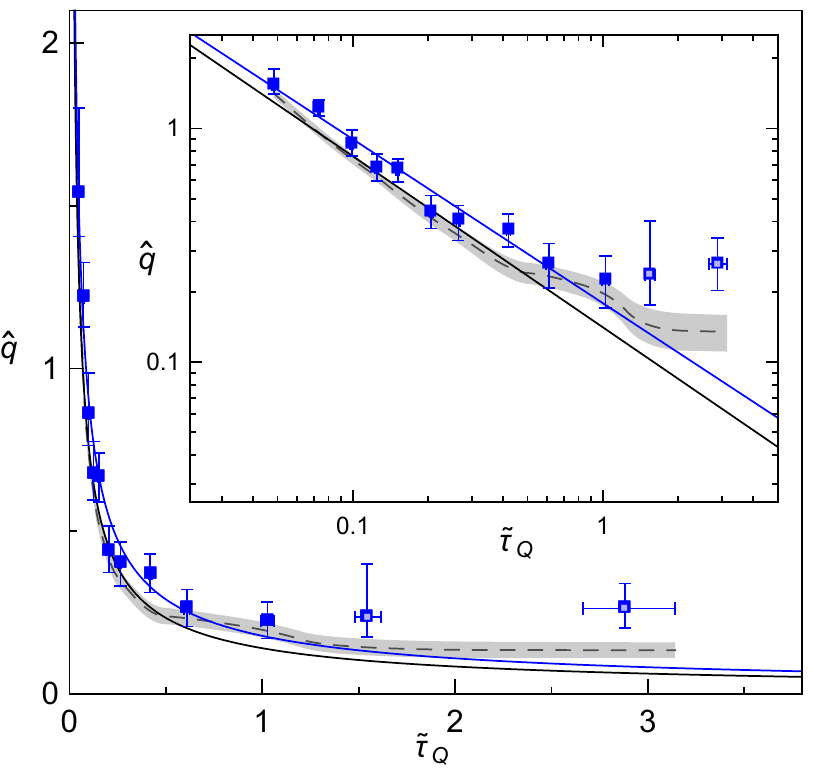}
    \caption{\textbf{Data set \#1, $\mathbf{\Delta\rho_0}$ threshold}. Plot range: 0.048 $ < \tilde{\tau}_Q < $ 2.88. Fit range: 0.048 $ < \tilde{\tau}_Q < $ 1.03. Critical exponents from fits: data: -0.70(6), simulations: -0.65(3).
    }
\label{fig:0127qhatdeltarho0}
	\end{minipage}
\end{figure}

\begin{figure}[h]
\centering
	\begin{minipage}{3.25in}
		 \includegraphics[scale=1]{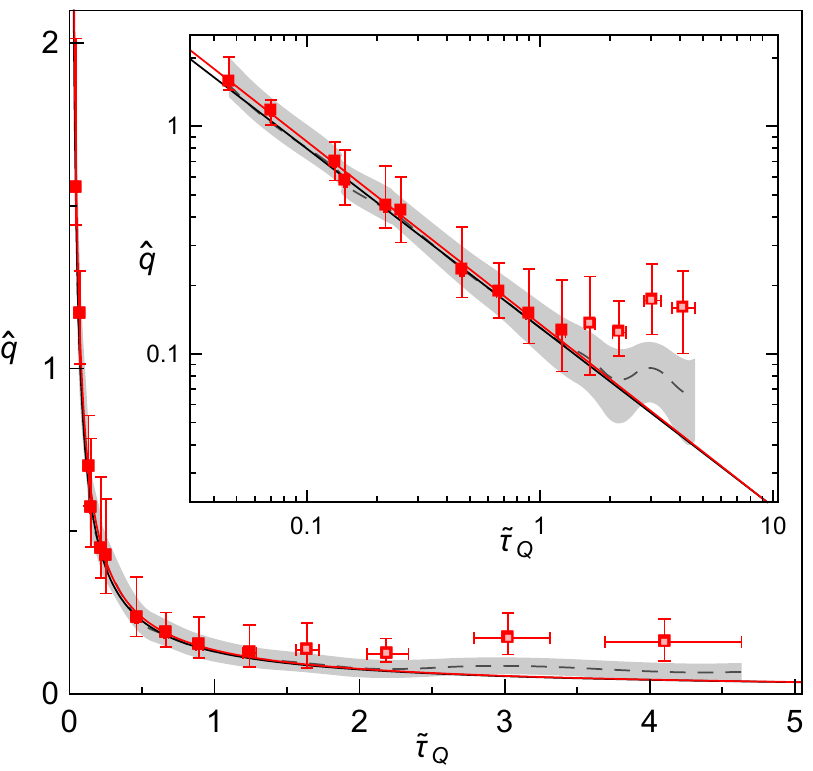}		
    \caption{\textbf{Data set \#2, $\mathbf{\rho_0}$ threshold}. Plot range: 0.046 $ < \tilde{\tau}_Q < $ 4.1. Fit range: 0.048 $ < \tilde{\tau}_Q < $ 1.24. Critical exponents from fits: data: -0.80(8), simulations: -0.79(7).
    }
\label{fig:0307qhatrho0}
	\end{minipage}
\hfill
	\begin{minipage}{3.25in}
		 \includegraphics[scale=1]{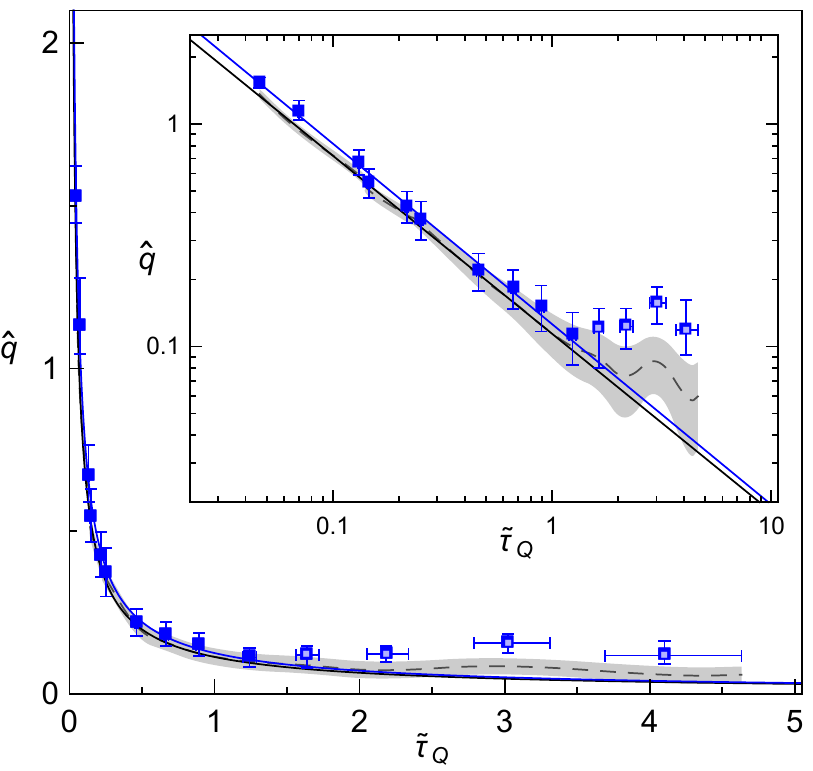}		
    \caption{\textbf{Data set \#2, $\mathbf{\Delta\rho_0}$ threshold}. Plot range: 0.046 $ < \tilde{\tau}_Q < $ 4.1. Fit range: 0.048 $ < \tilde{\tau}_Q < $ 1.24. Critical exponents from fits: data: -0.81(4), simulations: -0.80(3).
    }
\label{fig:0307qhatdeltarho}
\end{minipage}
\end{figure}

\begin{figure}[h]
\centering
	\begin{minipage}{3.25in}
		 \includegraphics[scale=1]{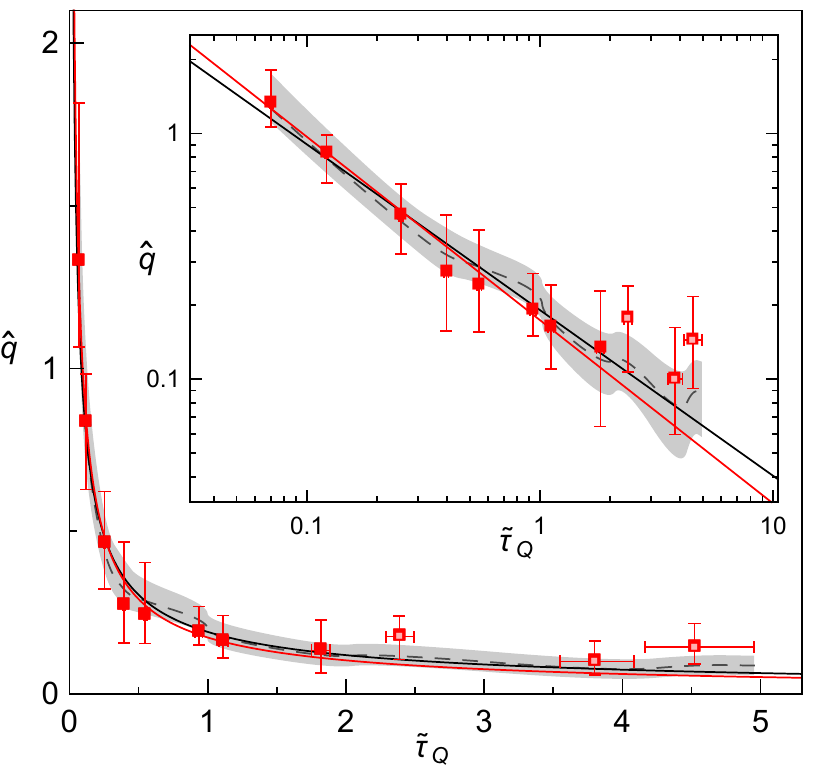}
    \caption{\textbf{Data set \#3, $\mathbf{\rho_0}$ threshold}. Plot range: 0.07 $ < \tilde{\tau}_Q < $ 4.52. Fit range: 0.07 $ < \tilde{\tau}_Q < $ 1.81. Critical exponents from fits: data: -0.75(11), simulations: -0.67(8).
    }
\label{fig:1116qhatrho0}
	\end{minipage}
\hfill
	\begin{minipage}{3.25in}
		 \includegraphics[scale=1]{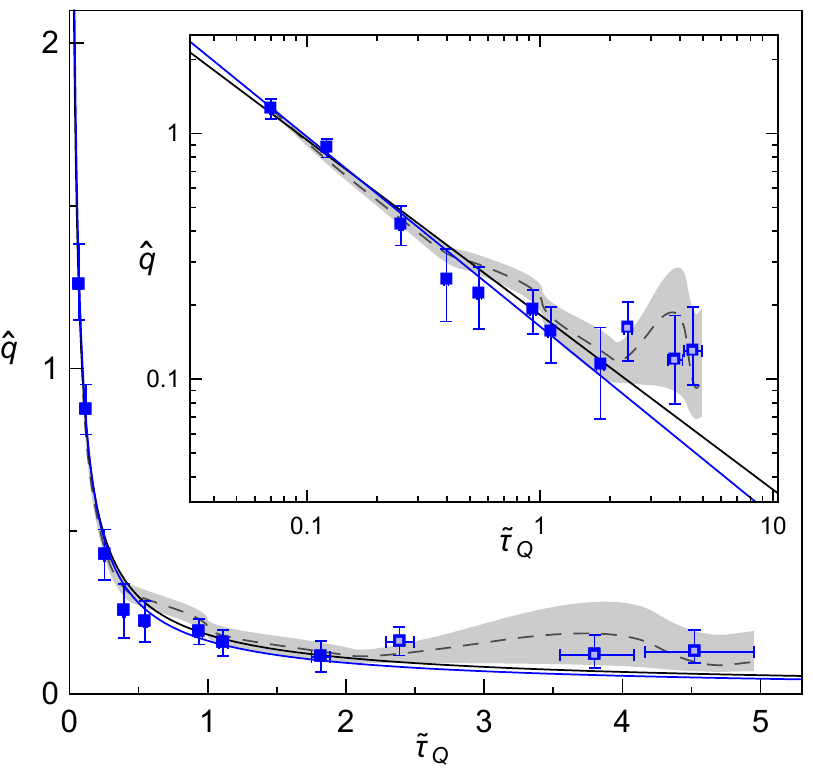}	
    \caption{\textbf{Data set \#3, $\mathbf{\Delta\rho_0}$ threshold}. Plot range: 0.07 $ < \tilde{\tau}_Q < $ 4.52. Fit range: 0.07 $ < \tilde{\tau}_Q < $ 1.81. Critical exponents from fits: data: -0.77(6), simulations: -0.70(3).
    }
\label{fig:1116qhatdeltarho0}
	\end{minipage}
\end{figure}

\begin{figure}[h]
\centering
	\begin{minipage}{3.25in}
		 \includegraphics[scale=1]{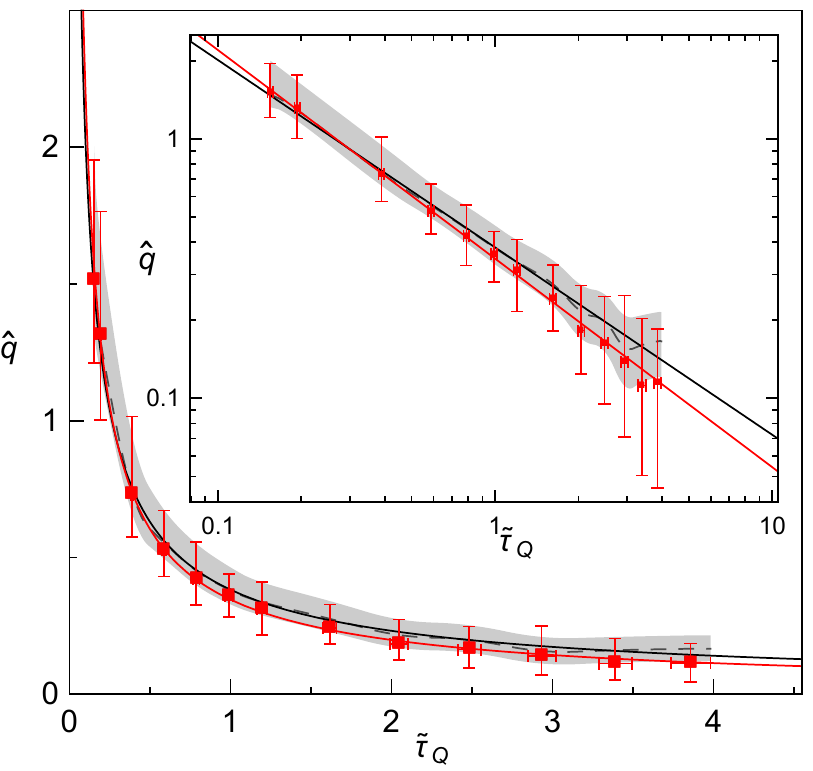}
    \caption{\textbf{Data set \#4, $\mathbf{\rho_0}$ threshold}. Plot range: 0.15 $ < \tilde{\tau}_Q < $ 3.86. Fit range: 0.15 $ < \tilde{\tau}_Q < $ 3.86. Critical exponents from fits: data: -0.80(10), simulations: -0.74(7).
    }
\label{fig:1216qhatrho0}
	\end{minipage}
\hfill
	\begin{minipage}{3.25in}
		 \includegraphics[scale=1]{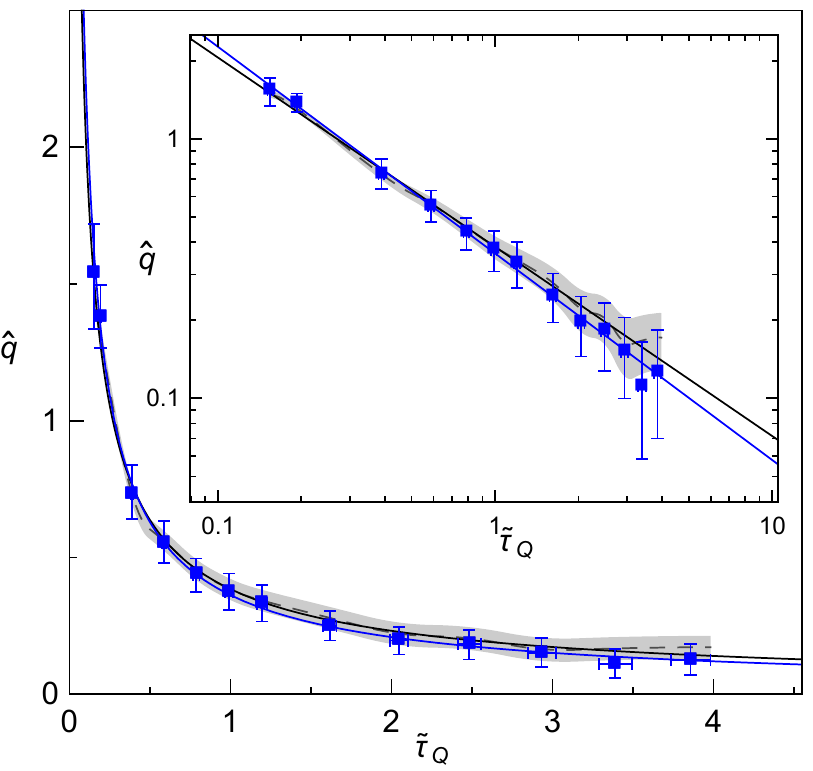}
    \caption{\textbf{Data set \#4, $\mathbf{\Delta\rho_0}$ threshold}. Plot range: 0.15 $ < \tilde{\tau}_Q < $ 3.86. Fit range: 0.15 $ < \tilde{\tau}_Q < $ 3.86. Critical exponents from fits: data: -0.80(5), simulations: -0.73(7).
    }
\label{fig:1216qhatdeltarho0}
	\end{minipage}
\end{figure}

\onecolumngrid
\newpage
\twocolumngrid

\onecolumngrid
\newpage
\twocolumngrid


\bibliographystyle{apsrev4-1}

\end{document}